\documentclass{scrartcl}

\usepackage[utf8]{inputenc}
\usepackage[T1]{fontenc}
\usepackage{authblk}

\usepackage{graphicx}
\usepackage{amsmath,amssymb,amsthm}
\usepackage{esint} 
\usepackage{enumitem,color}
\usepackage{comment}
\usepackage{hyperref}
\usepackage{wasysym}
\usepackage{relsize}

\usepackage{xstring}

\newtheorem{thm}{Theorem}
\newtheorem{prop}[thm]{Claim}
\newtheorem{trueprop}[thm]{Proposition}

\newtheorem{cor}[thm]{Corollary}
\theoremstyle{definition}

\theoremstyle{remark}
\newtheorem{rem}[thm]{Remark}

\newcommand{\fig}[2]{\includegraphics[width=#1\textwidth]{#2}}

\newcommand{\R}{\mathbb{R}}

\newcommand{\A}{\boldsymbol{A}}
\newcommand{\B}{\boldsymbol{B}}
\newcommand{\C}{\boldsymbol{C}}
\newcommand{\X}{\boldsymbol{X}}

\newcommand{\btwo}{\boldsymbol{b_2}}
\definecolor{darkolivegreen}{rgb}{0.33, 0.42, 0.18}
\definecolor{britishracinggreen}{rgb}{0.0, 0.26, 0.15}
\definecolor{ao(english)}{rgb}{0.0, 0.5, 0.0}

\title{Hamiltonian cycles on bicolored random planar maps}
\author[ ]{Bertrand Duplantier\thanks{bertrand.duplantier@ipht.fr}}
\author[ ] {Olivier Golinelli\thanks{olivier.golinelli@ipht.fr}}
\author[ ]{Emmanuel Guitter\thanks{emmanuel.guitter@ipht.fr}}
\affil[ ]{{\normalsize Université Paris-Saclay, CEA, CNRS, Institut de Physique
    Théorique\\ 91191~Gif-sur-Yvette, \textsc{France}}}

\date{May 3, 2023}

\begin{document}
\maketitle 
\begin{abstract}
We study the statistics of Hamiltonian cycles on various families of \emph{bicolored} random planar maps (with the spherical topology). These families fall into two groups
corresponding to two distinct universality classes with respective central charges $c=-1$ and $c=-2$. The first group includes generic \emph{$p$-regular} maps with vertices
of fixed valency $p\geq 3$, whereas the second group comprises maps with vertices of \emph{mixed} valencies, and the so-called \emph{rigid} case
of $2q$-regular maps ($q\geq 2$) for which, at each vertex, the unvisited edges are equally distributed on both sides of the cycle. We predict for each class its universal configuration exponent $\gamma$, as well as a new universal critical exponent $\nu$ characterizing the number of long-distance
contacts along the Hamiltonian cycle. These exponents are theoretically obtained by using the Knizhnik, Polyakov and Zamolodchikov (KPZ) relations, with
the appropriate values of the central charge, applied, in the case of $\nu$, to the corresponding critical exponent on regular (hexagonal or square) lattices. These predictions
 are numerically confirmed 
by analyzing exact enumeration results for $p$-regular maps with $p=3,4,\ldots,7$, and for maps with mixed valencies $(2,3)$, $(2,4)$ and $(3,4)$.
 \end{abstract}

\section{Introduction}
\label{sec:intro}

A \emph{planar map} is a connected graph embedded in the two-dimensional sphere without edge crossings, and considered up to homeomorphisms. 
A map is characterized by its vertices, its edges and its faces which all have the topology of the disk.  In this paper, the \emph{size} of a map is defined 
as its number of vertices.  A planar map is \emph{bicolored} if its vertices are colored in black and white so that edges connect only vertices of different 
colors. A \emph{Hamiltonian cycle} is a closed self-avoiding path drawn along the edges of the map that visits all the vertices of the map. This paper 
addresses the combinatorial problem of enumerating Hamiltonian cycles on various families of bicolored planar maps. Note that the length of a Hamiltonian cycle
on a bicolored map is necessarily an even integer and we shall denote it by $2N$, which is also the size of the underlying map. 

For a given family of bicolored 
planar maps, we will denote by $z_N$ the number of configurations of such maps with size $2N$, equipped with a Hamiltonian cycle and with a \emph{marked visited edge}. 
The quantity $z_N$ will be referred to as the partition function\footnote{This should more precisely be called a ``rooted partition function'' since we decided to mark
an edge of the configuration. This marking is convenient as it prevents configurations from having internal symmetries.} of the model at hand. At large $N$, we expect the asymptotic behavior 
\begin{equation}
\label{eq:asympz} 
z_N \underset{N\to \infty}{\sim} \varkappa \frac{\mu^{2N}}{N^{2-\gamma}}\ ,
\end{equation}
where $\varkappa$ and $\mu$ depend on the precise family of maps we are dealing with, while the \emph{configuration exponent} $\gamma$
has a more universal nature: as we shall see, only two possible values of $\gamma$ will be encountered, and it is precisely the aim of
this paper to understand when and why one or the other value is observed. 

As was done in \cite{D2G2} in the case of \emph{bicubic} maps (i.e., bicolored maps with only 3-valent vertices), we shall argue in the next section that 
the asymptotic properties of our Hamiltonian cycles on planar bicolored maps may be captured by viewing the problem as the coupling to gravity of a particular critical statistical 
model described by a conformal field theory (CFT), whose central charge $c$ may itself be deduced from a height reformulation of the problem.  
More precisely, it is known from the celebrated Knizhnik Polyakov Zamolodchikov (KPZ) formulas \cite{KPZ88,FD88,DK89} that the coupling to gravity of a CFT 
with central charge $c\leq 1$ corresponds to a fixed size (rooted) partition function $z_N$ with asymptotics  \eqref{eq:asympz} where (in the planar case considered in this paper):
\begin{equation}
\label{eq:KPZ}
\gamma=\gamma(c):=\frac{1}{12}\left(c-1-\sqrt{(1-c)(25-c)}\right)\ .
\end{equation}
As it will appear, the various families of bicolored maps that we shall study fall into two categories: when equipped with Hamiltonian cycles, some families 
will correspond to a CFT with central charge $c=-1$ and therefore exhibit a configuration exponent $\gamma=\gamma(-1)=-(1+\sqrt{13})/6$, while
the other families correspond to a CFT with central charge $c=-2$, leading to a configuration exponent $\gamma=\gamma(-2)=-1$. 

In our discussion, it will prove useful to extend our Hamiltonian cycle problem to that, more general, of \emph{fully packed loops} (FPL)
on planar bicolored maps. A fully packed loop configuration on a given map is defined as a set of self- and mutually avoiding loops drawn on
the edges of the map such that every vertex is visited by a loop. The lengths of all loops are again even, with total length equal to $2N$, and we finally attach 
a weight $n$ to each loop: this defines the so-called FPL$(n)$ model on the family of bicolored maps at hand. The case of Hamiltonian cycles may
be recovered from the $n\to 0$ limit of the FPL$(n)$ model.
 
\begin{rem}
The FPL$(n)$ model itself may be viewed as a particular critical point of the two-dimensional O$(n)$ model. Recall that this latter model
describes configurations of self- and mutually avoiding loops with a weight $n$ per loop and a \emph{fugacity $x$ per vertex visited  by a loop}.
The FPL$(n)$ model is thus recovered within the O$(n)$ model framework by letting $x\to \infty$ so that all vertices be visited
by a loop. As we shall recall later, the FPL$(n)$ model is intimately linked to the \emph{dense critical phase} of the O$(n)$ model.
\end{rem}
\begin{rem}
Let us insist on the fact that \emph{all the maps that we consider in this paper are bicolored}. As explained in \cite{GKN99,D2G2}, this coloring constraint 
is crucial when it comes to identifying the central charge of the associated CFT. We will comment on this in Remark~\ref{rem:monocol} below. 
\end{rem}

We also address the question 
of long-distance contacts within Hamiltonian cycles on random planar maps. Marking two points at distance $N$ along a cycle splits the latter into two equal parts, 
and defines a set of \emph{contact links}, i.e., edges that are incident to \emph{both} parts of the cycle; these contact links can be seen as connected by a \emph{dual contact cycle} on the dual map. 
Their average number scales as $N^\nu$, with a new exponent $\nu$ depending
on the underlying map family. The values of $\nu$ are predicted theoretically by using, for the proper value of the central charge $c$, the KPZ formula applied to a similar exponent on regular (hexagonal or square) lattices, 
which is (half) the Hausdorff dimension of contacts within a loop of the regular $\mathrm{FPL}(n\to 0)$ model.    
As we shall see, in the scaling limit, a bicolored random planar map equipped with a Hamiltonian cycle is expected to converge to a \emph{Liouville quantum gravity (LQG) sphere} \cite{DMS21}, decorated by an  \emph{independent (space-filling) whole-plane Schramm-Loewner evolution} $\mathrm{SLE}_8$ \cite{OS}, 
the dual contact cycle itself converging to a dual \emph{whole-plane} $\mathrm{SLE}_2$. The LQG parameter is $\gamma_{\scriptscriptstyle{\mathrm{L}}}=2/\sqrt{1-\gamma}$, with  either $\gamma=\gamma(c=-1)$ or $\gamma=\gamma(c=-2)$, depending on the chosen map's family.  
These predictions are in the same spirit as those made in Refs. \cite{D2G2,BGS22,BGP22}.

\medskip
The paper is organized as follows: 
Section~\ref{sec:pregular} discusses Hamiltonian cycles on bicolored planar maps whose \emph{all vertices have the same
valency} and gives our prediction for the configuration exponent $\gamma$ in this case. Section~\ref{sec:mixted} deals on the contrary with 
the case of maps having several allowed vertex valencies, leading to another value of $\gamma$. The predictions of these two sections
are verified numerically in Section~\ref{sec:numerics} by analyzing exact enumeration results for maps of finite sizes. Section~\ref{sec:rigid}
introduces the notion of \emph{rigid} Hamiltonian cycles and predicts a configuration exponent different from that of the non-rigid case.
This result is confirmed by the derivation of exact expressions for $z_N$ for arbitrary $N$. Section~\ref{sec:middle} addresses the question 
of long-distance contacts within Hamiltonian cycles, whose average number scales as $N^\nu$ with the exponent $\nu$ depending
on the underlying map family. Two possible values of $\nu$ are predicted theoretically and then checked numerically in Section~\ref{sec:numcontacts}. We conclude
with a few remarks in Section~\ref{sec:conclusion}.

\section{The case of $\boldsymbol{p}$-regular bicolored maps}
\label{sec:pregular}
Recall that a \emph{$p$-regular map} is a map whose all vertices have valency $p$. This section is devoted to the enumeration
of Hamiltonian cycles on $p$-regular bicolored planar maps for a fixed integer $p\geq 3$. It includes in particular the case of bicubic maps 
($p=3$) studied in \cite{D2G2,GKN99}.

\begin{figure}
  \centering
  \fig{.6}{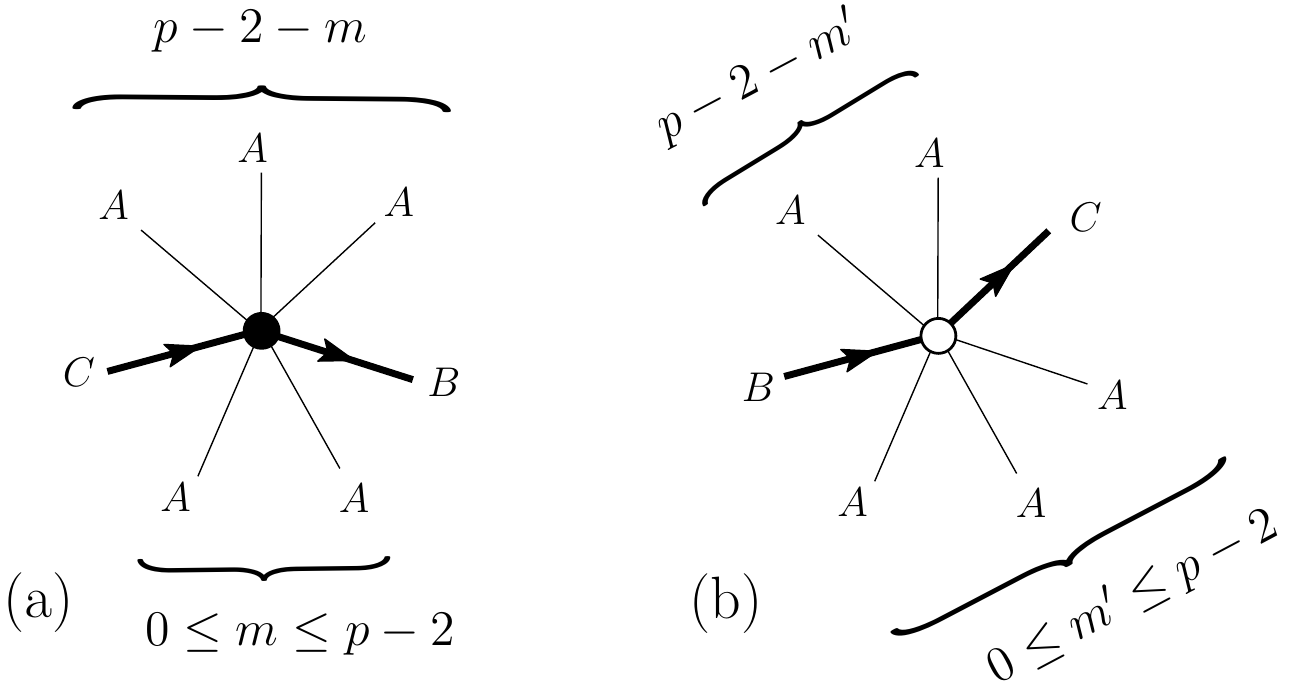}
   \caption{\small Example of the edge environment of (a) a black vertex and (b) a white vertex. Each vertex is surrounded by a $B$-edge, a $C$-edge
   and a total of $(p-2)$ $A$-edges (here $p=7$, $m=2$ and $m'=3$).  }
  \label{fig:environments}
\end{figure}

From now on, we therefore assume that $p$ takes a fixed value and we start by considering
the FPL$(2)$ model on $p$-regular bicolored planar maps. Assigning the weight $n=2$ per loop amounts equivalently to having unweighted
\emph{oriented} loops (the weight $2$ arising then from the $2$ possible orientations for each loop). This allows us to define three types
of edges (see Figure~\ref{fig:environments}): the unvisited edges, called $A$-edges, the edges visited by a loop whose orientation points toward their white incident vertex, 
which we call $B$-edges, and finally the edges visited by a loop whose orientation points toward their black incident vertex, 
which we call $C$-edges. The configuration of edges around a black vertex is then that of Figure~\ref{fig:environments}-(a) with an ingoing $C$-edge, 
an outgoing $B$-edge and a total of $p-2$ unvisited $A$-edges which are distributed in all possible ways on both sides of the loop. Similarly,
the configuration of edges around a white vertex is that of Figure~\ref{fig:environments}-(b) with now an ingoing $B$-edge, 
an outgoing $C$-edge and $p-2$ unvisited $A$-edges. 

We may now transform the FPL$(2)$ model into a \emph{$d$-component height model} by
assigning a height $\X\in\R^d$ to each face of the map, whose variation $\Delta\X$ between adjacent 
faces depends on the nature of the edge between them according to the rules of Figure~\ref{fig:heights}: 
we demand that $\Delta \X = \A$ (resp.\ $\B$, $\C$) if the crossed edge is of type $A$ (resp.\ $B$, $C$) and traversed with its incident white vertex on the right. 
To guarantee that the height is well defined across the whole map, we have to ensure that we recover the same value of $\X$ after making a complete 
turn around any vertex of the map. This requires (see Figure~\ref{fig:heights}) the necessary and sufficient condition:
\begin{figure}
  \centering
  \fig{.7}{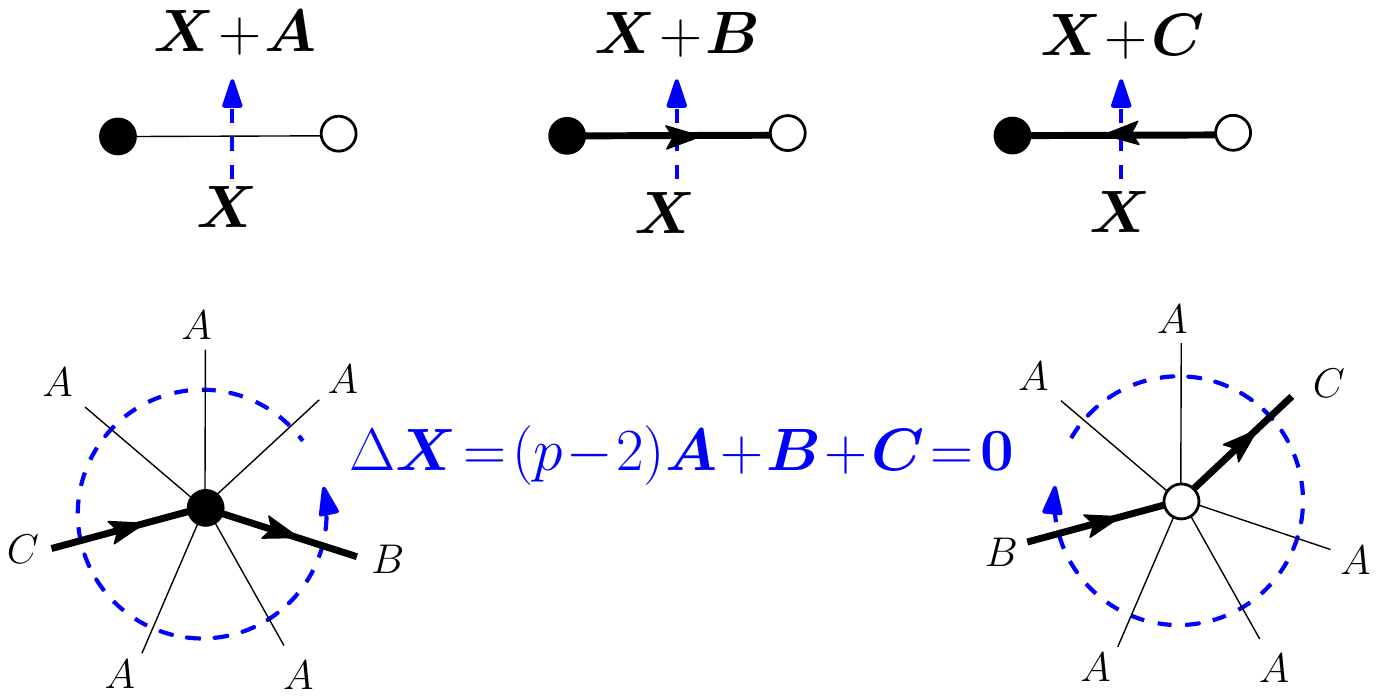}
   \caption{\small Top: rules for the variation in the height variable $\X$ when crossing an edge of the map. Bottom: making a complete turn counterclockwise (resp.\ clockwise) around 
   a black (resp. white) vertex results in a height variation $\Delta \X=(p-2)\A+\B+\C$ which for consistency must be taken equal to $\boldsymbol{0}$.}
  \label{fig:heights}
\end{figure}
\begin{equation}
(p-2)\A+\B+\C=\boldsymbol{0}
\label{eq:sumrule}
\end{equation}
which, de facto, implies that $\X$ lives in the $(\B,\C)$ \emph{two-dimensional} plane. For definiteness, we choose
$d=2$ with $\B$ and $\C$ two unit vectors of $\R^2$ satisfying, say $\B\cdot\C=-1/2$. In particular, the property $|\B|=|\C|$ implies that $(\B-\C)\cdot(\B+\C)=0$ hence,
if we define 
\begin{equation}
\btwo:=\B-\C\ ,
\end{equation}
we deduce from \eqref{eq:sumrule} that $\A\cdot\btwo=0$ and a natural convention consists in expressing 
our two-component height variable $\X$ in the \emph{orthogonal} basis $(\A,\btwo)$. In the continuous limit, we expect that the FPL$(2)$ model is
therefore described by the coupling to gravity of a two-dimensional CFT involving a \emph{two-component vector field} $\boldsymbol{\Psi}=\psi_1 \A+\psi_2 \btwo$ 
(i.e., with components both along $\A$ and along $\btwo$) measuring locally the ``coarse grained'' averaged value $\boldsymbol{\Psi}  = \langle \X \rangle$ and 
governed by a free field action for both $\psi_1$ and $\psi_2$, see \cite{D2G2}. 
We deduce the following:
\begin{prop}
\label{claim1}
The FPL$(2)$ model on $p$-regular bicolored planar maps is described by the coupling to gravity of a CFT
with central charge 
\begin{equation}
c=2\ .
\end{equation}
\end{prop} 

We now wish to understand how this result is modified if we give an arbitrary weight $n$ to each loop, hence consider the FPL$(n)$ model
on  $p$-regular bicolored planar maps. The case $p=3$ of bicubic maps was discussed in details in \cite{D2G2}. There, the underlying CFT
is identified as that describing the FPL$(n)$ model on the \emph{honeycomb, i.e., hexagonal lattice} (which is $3$-regular and can be bicolored canonically),
a model well studied in \cite{Re91,BN94,BSY94,KdGN96} by Bethe Ansatz or Coulomb gas techniques.  For this lattice model,
the passage from $n=2$ to an arbitrary $n\in [-2,2]$ modifies in the continuous limit the Gaussian free field action by adding a term which couples the component
 $\psi_2$ of the two-component field $\boldsymbol{\Psi}$ to the local intrinsic curvature of the underlying surface, while
 the action for the component $\psi_1$ remains that of a free field. For $n\in [-2,2]$, the net result 
 is a shift of the central charge from $c=c_{\mathrm{fpl}}(2)=2$ to a lower value $c=c_{\mathrm {fpl}}(n)$ \cite{BN94} whose expression is recalled just below. 

Since the FPL$(2)$ model on $p$-regular bicolored planar maps has the same two-component field description for any arbitrary integer $p\geq 3$, 
we expect that the passage from $n=2$ to an arbitrary $n\in [-2,2]$ induces the very same lowering of the central charge. This leads us to express the following statement:
\begin{prop}
\label{claim2}
For $-2\leq n\leq 2$, the FPL$(n)$ model on $p$-regular bicolored planar maps is, for arbitrary $p\geq 3$, described by the coupling to gravity of a CFT
with central charge 
\begin{equation}
\label{eq:valcgn}
c=c_{\mathrm{fpl}}(n):=2-6\frac{(1-g)^2}{g}\quad \mathrm{where}\  n=-2\cos(\pi\, g)\ \mathrm{with}\ 0\leq g\leq 1 \ .
\end{equation}
\end{prop} 
In the $n\to0$ limit, we deduce in particular:
\begin{cor}
\label{claim3}
The model of Hamiltonian cycles on $p$-regular bicolored planar maps is described by the coupling to gravity of a CFT
with central charge 
\begin{equation}
c=c_{\mathrm{fpl}}(0)=-1\ .
\end{equation}
In particular, using KPZ \eqref{eq:KPZ}, the partition function $z_N$ of Hamiltonian cycles on $p$-regular bicolored planar maps of size $2N$ has the 
asymptotic behavior \eqref{eq:asympz} with
\begin{equation}\label{eq:gamma-1}
\gamma=\gamma(-1)=-\frac{1+\sqrt{13}}{6}\ .
\end{equation}
\end{cor} 
This extends the conjecture of \cite{GKN99} (see also \cite{DFG05,D2G2}) for $p=3$ to an arbitrary value of the 
integer $p\geq 3$. 

\begin{figure}
  \centering
  \fig{.6}{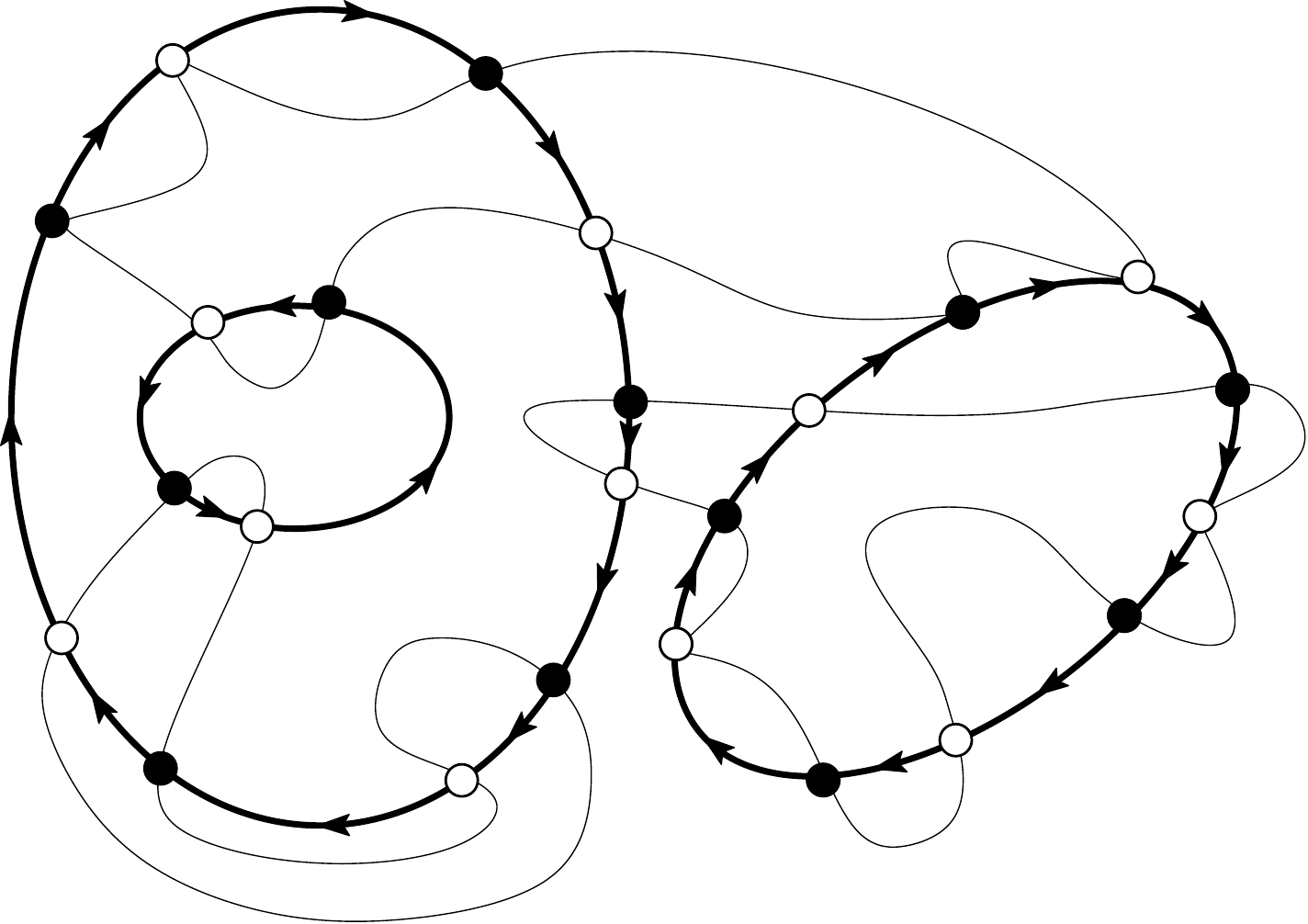}
   \caption{\small An example of a $4$-regular bicolored planar map equipped with a set of fully packed oriented loops (thick lines). The unvisited edges (thin lines) 
   automatically form a complementary set of fully packed unoriented loops on the map.}
  \label{fig:pfour}
\end{figure}
It is interesting to remark that for $p=4$ we may arrive at the statements of Claim \ref{claim2}
and Corollary \ref{claim3} by a different route as follows. For $p=4$, each vertex is incident to exactly $2$ unvisited edges: the unvisited edges thus naturally form loops visiting all the vertices of the bicolored map, see Figure~\ref{fig:pfour}.
We therefore have by construction two complementary systems of fully packed loops: the original fully packed loops which receive a weight $n_1=n$ and
the loops formed by the unvisited edges which receive the neutral weight $n_2=1$.  
The FPL$(n)$ model on $4$-regular bicolored planar maps may therefore be viewed as a particular instance of 
the coupling to gravity of the so-called FPL$^2(n_1,n_2)$ model, which involves two complementary fully packed loop systems with respective weights $n_1$ and $n_2$ on the square lattice (which is $4$-regular and can be bicolored canonically).
The FPL$^2(n_1,n_2)$ model on this lattice was studied in details in \cite{BBNY96,JK98,DCN04,JZJ09} by Coulomb gas and Bethe Ansatz techniques. Its central
charge was found to equal $c_{\mathrm{fpl}^2}(n_1,n_2)=3-6(1-g_1)^2/g_1-6(1-g_2)^2/g_2$ where, for $i=1,2$, 
$n_i=-2\cos(\pi\, g_i)$ with $0\leq g_i\leq 1$. Taking $g_1=g $ as in \eqref{eq:valcgn} and $g_2=2/3$ so that $(n_1,n_2)=(n,1)$, 
we recover the value $c_{\mathrm{fpl}^2}(n,1)=2-6(1-g)^2/g=c_{\mathrm{fpl}}(n)$ as in Claim \ref{claim2} and Corollary \ref{claim3}  .
 
Note that, for $p\geq 5$, we can no longer rely on hypothetical results for a fully packed loop model on some regular lattice, 
since there exists no such bicolored regular lattice with $p$-valent vertices only\footnote{For $p=6$, a natural candidate with
only 6-regular vertices is the triangular lattice but this lattice is not bicolorable.}. Moreover, for $p\geq 5$, there is no canonical 
way to arrange the unvisited edges into loops, would it be only for a subset of these unvisited edges.

\section{The case of bicolored maps with mixed valencies}
\label{sec:mixted}
In this section, we deal with planar maps whose vertices have valencies within the fixed set $\mathcal{S}=\{p_1,p_2,\ldots,p_k\}$ where 
$k\geq 2$  and where the integers $p_i$ satisfy $2\leq p_1<p_2<\cdots<p_k$. Such maps will be generically referred to  as 
\emph{maps with mixed valencies}. 
Again we are interested in evaluating the number of such bicolored maps equipped with a Hamiltonian cycle, or more generally a set
of fully packed loops with a weight $n$ per loop. Since the (self- and mutually-avoiding) loops visit all the vertices, the underlying maps 
have by construction an even size $2N$, with exactly $N$ black and $N$ white vertices.  The statistical 
ensemble that we consider is that \emph{with fixed $N$} and with a weight $w_i\in \R^{+}$ attached to each vertex with valency $p_i$.  
We insist here on the fact that the numbers $m_i$ of vertices of valency $p_i$ are not fixed individually but that their sum 
$m_1+m_2+\cdots+m_k=2N$ is fixed. We call $z_N$ the associated partition function with, as before, a marked visited edge. The partition
function $z_N$ depends implicitly on the set $\mathcal{S}$ and on the weights $w_i$. Note that, since $w_i>0$ for all $i\in \{1,\ldots,k\}$, we
expect the average number of vertices $\langle m_i\rangle=w_i \frac{\partial \ }{\partial w_i} \mathrm{Log}z_N$ to be of order $N$ for all $i$'s, i.e., extensive
for each valency $p_i$.

\medskip
As in the previous section, we start by studying the FPL$(2)$ model on our bicolored maps with mixed valencies and fixed size $2N$.
As before, the weight $2$ per loop can be realized by orienting the loops, and we may again describe alternatively the configurations by
a $d$-component height variable $\X\in \R^d$ defined from the loop content according to the rules of Figure~\ref{fig:heights}-top. 
Note that configurations where valencies belong only to a proper subset of $\mathcal{S}$ may appear. However, since
all weights $w_i$, $i\in \{1,2,\ldots,k\}$ have been chosen to be strictly positive, the asymptotic behavior of the partition function $z_N$ is exponentially dominated by
configurations where all valencies are \emph{macroscopically} present.
Considering two different valencies, say $p_{i_1}$ and $p_{i_2}$, we must, in order to have a well 
defined uni-valued height, impose simultaneously the two conditions $(p_{i_1}-2)\A+\B+\C=\boldsymbol{0}$ (necessary around a vertex of valency $p_{i_1}$)
and $(p_{i_2}-2)\A+\B+\C=\boldsymbol{0}$ (necessary around a vertex of valency $p_{i_2}$). Since we assumed $p_{i_1}\neq p_{i_2}$,
these two conditions imply
\begin{equation}
\A=\boldsymbol{0}\quad \mathrm{and} \quad \B+\C=\boldsymbol{0}\ .
\label{eq:Azero}
\end{equation}
This now implies that $\X$ stays colinear to $\B$, or equivalently to $\btwo:=\B-\C=2\B$.
In the continuous limit, we expect that the FPL$(2)$ model is
now described by the coupling to gravity of a two-dimensional CFT involving a \emph{one-component field} $\boldsymbol{\Psi}=\psi_2 \btwo$ 
(i.e., with a components along $\btwo$ only, so that we may in practice fix $d=1$) measuring as before the ``coarse grained''  averaged value $\boldsymbol{\Psi}  = \langle \X \rangle$ and 
governed by a Gaussian free field action.
This leads us to the following:
\begin{prop}
\label{claim4}
The FPL$(2)$ model on bicolored planar maps with mixed valencies is described by the coupling to gravity of a CFT
with central charge 
\begin{equation}
c=1\ .
\end{equation}
\end{prop} 
As for the case of arbitrary $n\in [-2,2]$, the action of the associated continuous CFT is again obtained
by adding to the free field action for $\psi_2$ a term which couples it to the local intrinsic curvature of the underlying surface.
Since there is no component $\psi_1$ anymore, the obtained central charge becomes equal to 
$c=c_{\mathrm{dense}}(n):=c_{\mathrm{fpl}}(n)-1$.
We arrive at:
\begin{prop}
\label{claim5}
For $-2\leq n\leq 2$, the FPL$(n)$ model on bicolored planar maps with mixed valencies is described by the coupling to gravity of a CFT
with central charge 
\begin{equation}\label{eq:valcgndense}
c=c_{\mathrm{dense}}(n):=1-6\frac{(1-g)^2}{g}\quad \mathrm{where}\  n=-2\cos(\pi\, g)\ \mathrm{with}\ 0\leq g\leq 1 \ .
\end{equation}
\end{prop} 
In the $n\to0$ limit, we deduce in particular:
\begin{cor}
\label{claim6}
The model of Hamiltonian cycles on bicolored planar maps with mixed valencies is described by the coupling to gravity of a CFT
with central charge 
\begin{equation}
c=c_{\mathrm{dense}}(0)=-2\ .
\end{equation}
In particular, using KPZ, the associated partition function $z_N$ has the 
asymptotic behavior \eqref{eq:asympz} with
\begin{equation}
\gamma=\gamma(-2)=-1\ .
\end{equation}
\end{cor} 
 
\begin{rem}
The denomination ``dense'' refers to the fact that the value $c_{\mathrm{dense}}(n)$ of the central charge is precisely that
associated with the two-dimensional O$(n)$ model in its dense critical phase, where the number of occupied vertices is macroscopic, with loops being no longer required to visit all the vertices (see Section~\ref{sec:scalinglimit} for a detailed discussion).
Here we recover this value even though, in our problem, loops by definition visit all vertices. The randomness due to the multiple choice of valencies somehow erases the full-packing constraint, which  corresponds to an unstable manifold in the parameter space of the $\mathrm{O}(n)$ model \cite{BN94}.
\end{rem}

\begin{rem}
\label{rem:monocol}
Note that a similar reduction in the central charge from $c_{\mathrm{fpl}}(n)$ to $c_{\mathrm{dense}}(n)=c_{\mathrm{fpl}}(n)-1$
would be observed for $p$-regular maps in the absence of the bicoloring constraint. Indeed, in that case, it is no longer possible
to distinguish the two sides of an $A$-edge (see Figure~\ref{fig:heights}-top), which forces one to set $\A=\boldsymbol{0}$
and thus $\B+\C=\boldsymbol{0}$ as in \eqref{eq:Azero}; see \cite{D2G2} for a detailed discussion in the $3$-regular map case. 
\end{rem}

\section{Numerical verification}
\label{sec:numerics}

\begin{figure}
  \centering
  \fig{1.}{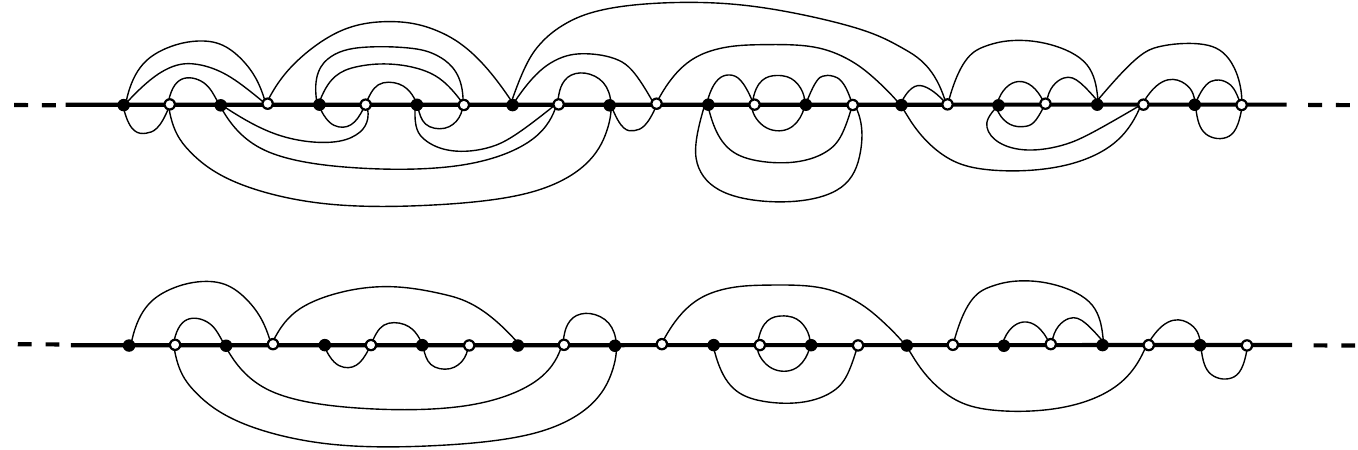}
   \caption{\small Representation of a Hamiltonian cycle (after opening its marked visited edge) as an infinite straight line with alternating black and white vertices,
   connected by non-crossing bicolored arches on both sides of the infinite line. Top: example in the $p$-regular case with $p=5$. Bottom: example in the
   case of mixed valencies $3$ and $4$, i.e., $k=2$ and $\mathcal{S}=\{3,4\}$.}
  \label{fig:largearches}
\end{figure}
\begin{figure}
  \centering
  \fig{.8}{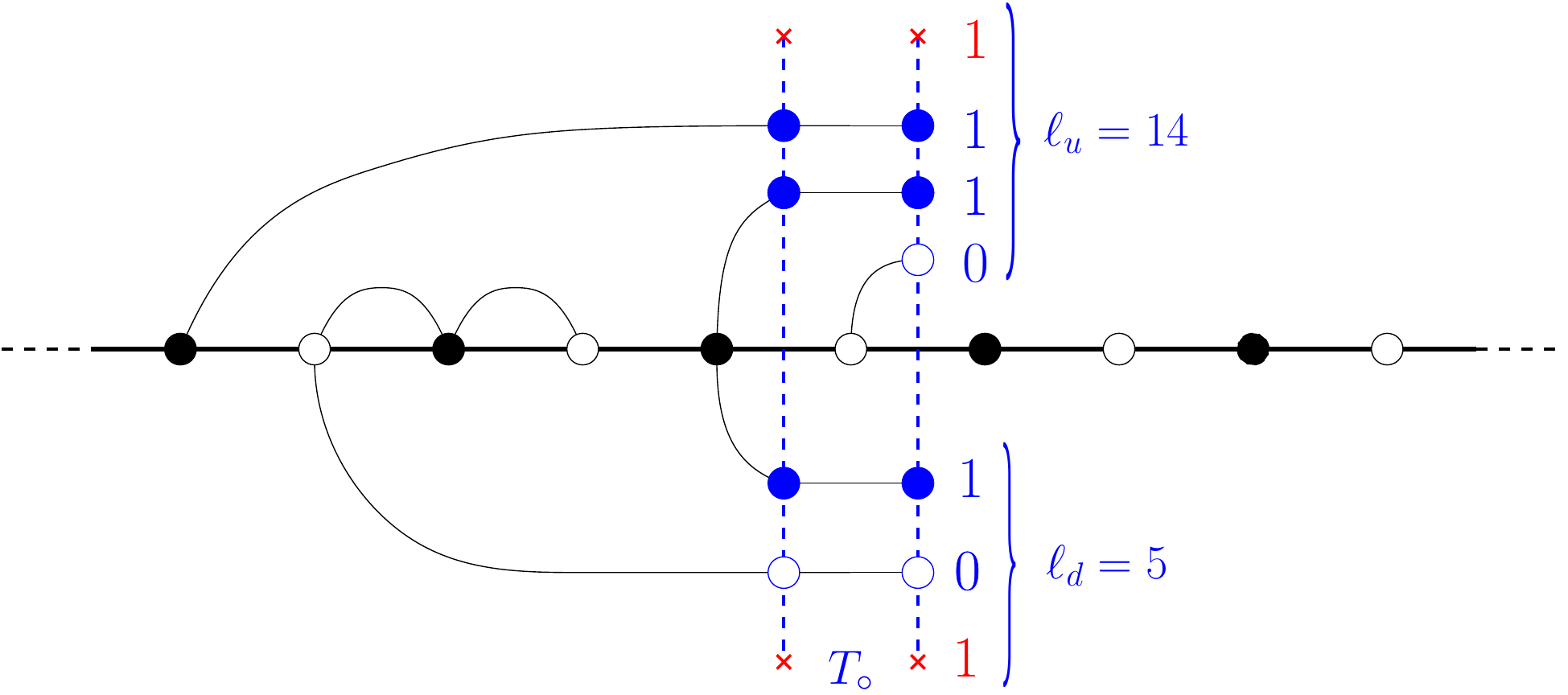}
   \caption{\small Illustration of the transfer matrix method in the case of mixed valencies with $\mathcal{S}=\{3,4\}$. Here we display one of the possible outcomes for the action of the elementary
   transfer matrix $T_\circ$ at the crossing of a white vertex. }
  \label{fig:TM}
\end{figure}

In order to verify the claims of Corollaries \ref{claim3} and \ref{claim6}, we performed a direct numerical enumeration of Hamiltonian cycles on various $p$-regular map families as well as
on various families of maps with mixed valencies. In all cases, by cutting the Hamiltonian cycle at the level of its marked visited edge and opening it
into a straight line, we obtain a configuration of the form of that in Figure~\ref{fig:largearches}, with an infinite line carrying $2N$ alternating black and white vertices. A vertex of valency $p_i$
leads to a total number $(p_i-2)$ of incident unvisited half-edges distributed in all possible ways on both sides of the infinite line. Finally, these half-edges are
connected in pairs so as to form a set of \emph{bicolored non-crossing arches}. To obtain the value of the number of possible configurations $z_N$ for a given map family,
we use a transfer matrix approach, generalizing that of \cite{D2G2}, in which the arch configurations are built from left to right along the straight
line of alternating black and white vertices. A transfer matrix state is described by the color sequence of those arches which have been opened but not yet closed, 
each arch inheriting the color of the vertex it originates from (see Figure~\ref{fig:TM}).
The upper arch color sequence is read from bottom to top and the lower one from top to bottom. A sequence of $s$ arches with 
colors $a_1,\ldots,a_s$ (where we choose $a_j=1$ for black and $0$ for white) is encoded by the integer $\ell=2^{s}+\sum_{j=1}^s a_j 2^{(j-1)}$ so that a transfer matrix intermediate state
is coded by two positive integers $\ell_{u}$ (upper sequence) and $\ell_{d}$ (lower sequence) and denoted as $|\ell_{u},\ell_{d}\rangle$. With these notations, the partition function
$z_N$ may be written as 
\begin{equation}
z_N=\langle1,1|( T_\circ T_\bullet)^N |1,1\rangle
\end{equation}
where $|1,1\rangle$ correspond to the empty configuration (the vacuum state) while $T_\bullet$ and $T_\circ$ are two elementary transfer matrices transferring the state respectively across
a black and a white vertex. 
Note that, for $N$ even, we may write
\begin{figure}
  \centering
  \fig{.8}{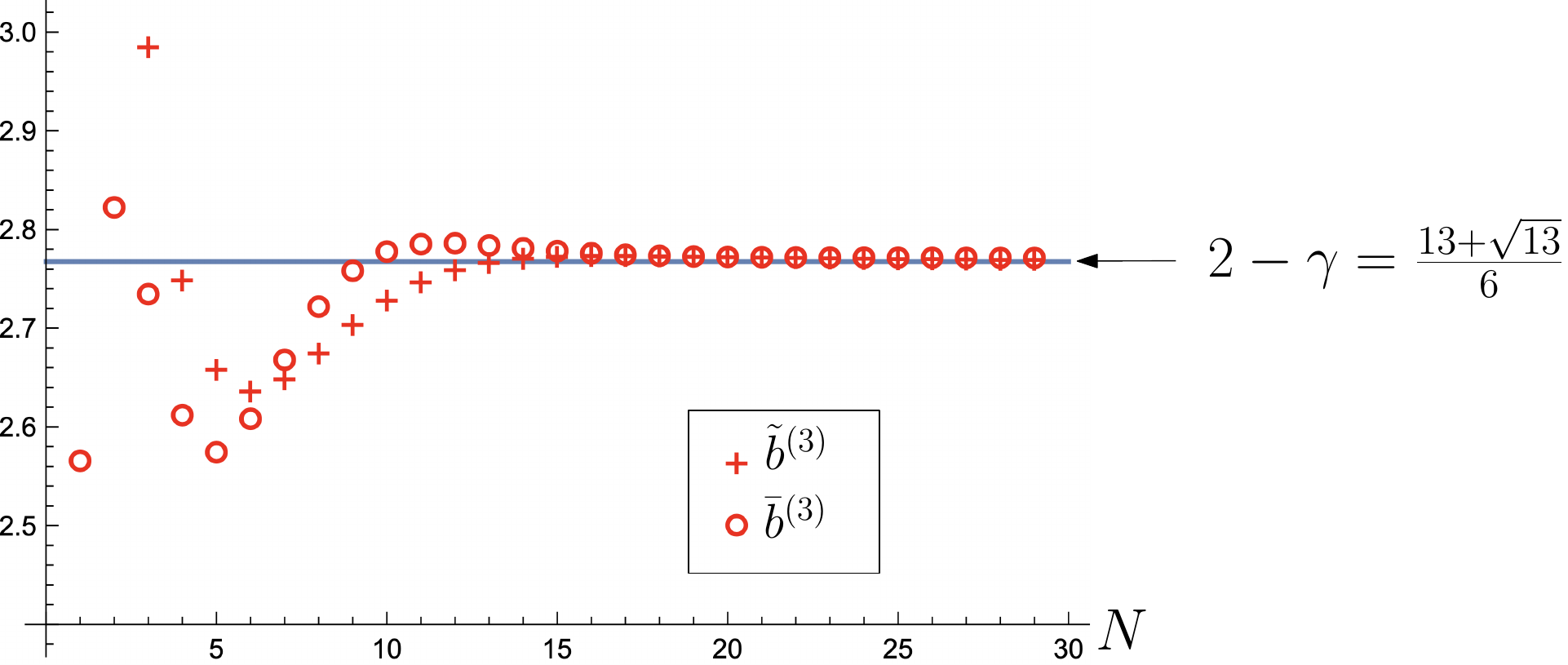}
   \caption{\small Estimates of $2-\gamma$ for Hamiltonian cycles on $3$-regular bicolored planar maps, as obtained
   from the associated accelerated series $\tilde{b}^{(3)}_N$ and $\bar{b}^{(3)}_N$ defined in \eqref{eq:bdef} and \eqref{eq:baccel}. These estimates confirm and extend the results
   of  \cite{GKN99} and \cite{D2G2}.}
  \label{fig:gamma3est}
\end{figure}
\begin{equation}
\begin{split}
z_N&=\sum_{\ell_u,\ell_d}\langle1,1|( T_\circ T_\bullet)^{N/2}|\ell_u,\ell_d\rangle \langle \ell_u,\ell_d| ( T_\circ T_\bullet)^{N/2}  |1,1\rangle\\
&= \sum_{\ell_u,\ell_d}\big(\langle \ell_u,\ell_d | ( T_\circ T_\bullet)^{N/2}  |1,1\rangle \big)^2\ ,
\end{split}
\label{eq:zNpair}
\end{equation}
where the sum is over the \emph{finite} number of reachable states after $N$ steps ($N/2$ of each color). Here we used 
the symmetry of the problem  under combined left-right reversal and black-white inversion of vertex colors.
Similarly, for $N$ odd, we have
\begin{equation}
\begin{split}
z_N&=\sum_{\ell_u,\ell_d}\langle1,1 |( T_\circ T_\bullet)^{(N-1)/2}T_\circ |\ell_u,\ell_d\rangle \langle \ell_u,\ell_d| T_\bullet ( T_\circ T_\bullet)^{(N-1)/2}  |1,1\rangle\\
&= \sum_{\ell_u,\ell_d}\big(\langle \ell_u,\ell_d | T_\bullet ( T_\circ T_\bullet)^{(N-1)/2}  |1,1\rangle \big)^2\ .
\end{split}
\end{equation}
We therefore see that, for both parities and for a total size $2N$ of the map configuration, we only have to perform the action of $N$ elementary transfer matrices. 

From $z_N$, we may obtain $\mu$ and $\gamma$ in \eqref{eq:asympz} as the limits of appropriate sequences: for instance the sequence
\begin{equation}
\label{eq:bdef}
 b_N:=N^2\, \mathrm{Log}\frac{z_{N+2}z_N}{(z_{N+1})^2}
 \end{equation}
tends to $2-\gamma$ for $N\to \infty$. We may therefore get an estimate for $\gamma$ from the value of $b_N$ for some finite, large enough, $N$.
To get a better estimate, we also have recourse to \emph{series acceleration methods}, involving sequences constructed from $b_N$ by recursive use 
of the finite difference operator $\Delta$ (defined by $(\Delta f)_N:=f_{N+1}-f_N$) and which converge faster to the same limit $2-\gamma$ as $N\to\infty$.
In practice, we use the two ``accelerated'' series $\tilde{b}_N$ and $\bar{b}_N$ defined as\footnote{The two series are defined so that their $N$'th element involves values of $z_M$
for $M$ up to $N+5$.}
\begin{equation}
\label{eq:baccel}
\begin{split}
\tilde{b}_N&:=\frac {1}{3!}(\Delta^{\!3}\, \hat{b})_N \quad \mathrm{with} \quad \hat{b}_N:=N^3 b_N\ ,\\
\bar{b}_N&:=b_{N+2}-2\frac{(\Delta b)_{N+2}(\Delta b)_{N+1}}{(\Delta^{2}\, b)_{N+1}}\ .\\
\end{split}
\end{equation} 

Appendix~\ref{app:numerics} presents our numerical results for the enumeration of $z_N$. More precisely, we deal with the following map families: 
\begin{itemize}
\item[-]{$p$-regular bicolored planar maps for $p=3,4,\ldots,7$};
\item[-]{bicolored planar maps with mixed valencies for $\mathcal{S}=\{2,3\},\{2,4\}$ with weights $w_2=w_3=w_4=1$ and
for $\mathcal{S}=\{3,4\}$ with $(w_3,w_4)=(1,1),(1,2)$ and $(2,1)$.}
\end{itemize}
\begin{figure}
  \centering
  \fig{.8}{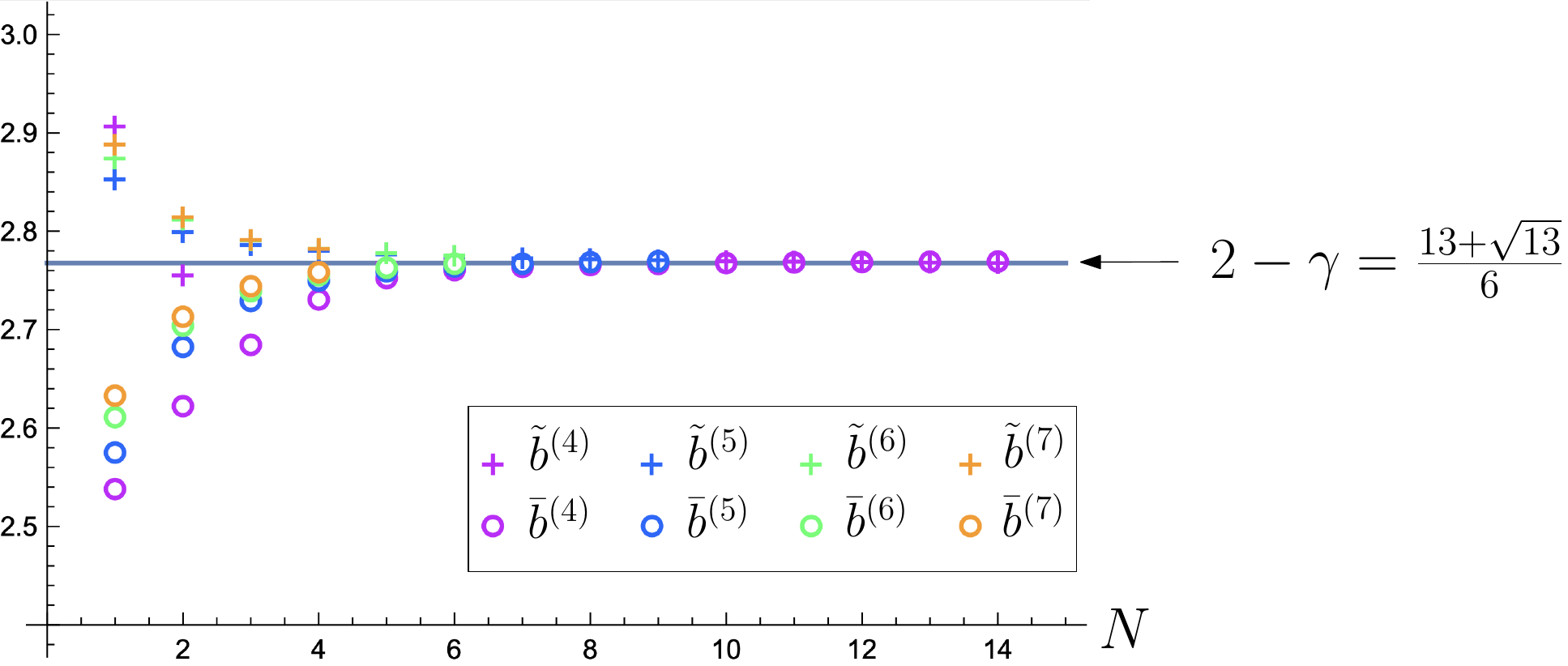}
   \caption{\small Estimates of $2-\gamma$ for Hamiltonian cycles on $p$-regular bicolored planar maps, as obtained
   from the associated accelerated series $\tilde{b}^{(p)}_N$ and $\bar{b}^{(p)}_N$ for $p=4,5,6$, and $7$. }
  \label{fig:gamma4-7est}
\end{figure}
From these values, we extract the estimates of $\mu^2$ listed in Table~\ref{table:valmu}.
\begin{table} [h]
    \centering
       \begin{tabular}{|c|c||c|c|}
        \hline
$p$-regular maps   &         $\mu^2$   & maps with mixed valencies   &   $\mu^2$    \\
        \hline
  $3$-regular   &   $10.113\pm0.001$    &  $\{2,3\} \quad (w_2=w_3=1)$  &  $16.204\pm0.001$  \\
  $4$-regular  &    $41.60\pm0.02$   &   $\{2,4\} \quad (w_2=w_4=1)$ &  $49.9\pm0.1$  \\
  $5$-regular   &     $117.0\pm0.2$  & $\{3,4\} \quad (w_3=w_4=1)$   & $86.02\pm0.05$  \\
 $6$-regular   &    $ 265.5\pm1.$  & $\{3,4\} \quad (w_3=1\ ,\  w_4=2)$  & $244.0\pm0.2$   \\
 $7$-regular   &     $522.8 \pm2.$  &  $\{3,4\} \quad (w_3=2\ ,\ w_4=1)$   &  $151.0\pm0.2$ \\  
       \hline
    \end{tabular}
    \caption{\small Estimated values of the exponential growth factor $\mu^2$.}
    \label{table:valmu}
\end{table}

Figures~\ref{fig:gamma3est} and \ref{fig:gamma4-7est} present our estimates of $2-\gamma$ for the $p$-regular bicolored planar maps with $p=3$
and $p=4$ to $7$ respectively (for each $p$, we denote by $b_N^{(p)}$ the associated series \eqref{eq:bdef}). 
These estimates are in perfect agreement with the expected value $\gamma=-(1+\sqrt{13})/6$ of Corollary~\ref{claim3}.

\begin{figure}
  \centering
  \fig{.8}{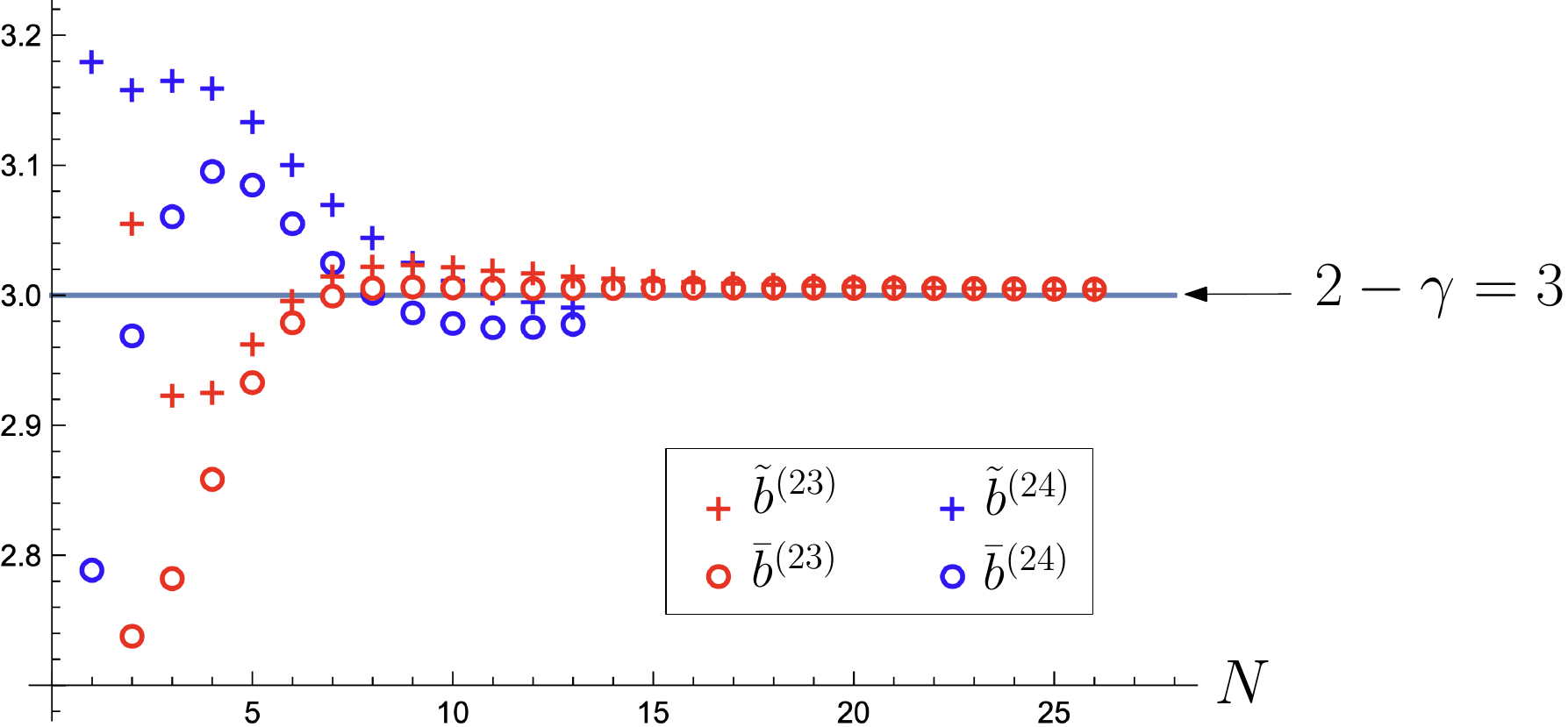}
   \caption{\small Estimates of $2-\gamma$ for Hamiltonian cycles on bicolored planar maps with mixed valencies
   in $\mathcal{S}=\{2,3\}$  (accelerated series $\tilde{b}^{(23)}_N$ and $\bar{b}^{(23)}_N$ with $w_2=w_3=1$) and in $\mathcal{S}=\{2,4\}$ 
    (accelerated series $\tilde{b}^{(24)}_N$ and $\bar{b}^{(24)}_N$ with $w_2=w_4=1$).}
  \label{fig:gamma23-24est}
\end{figure}

\begin{figure}
  \centering
  \fig{.8}{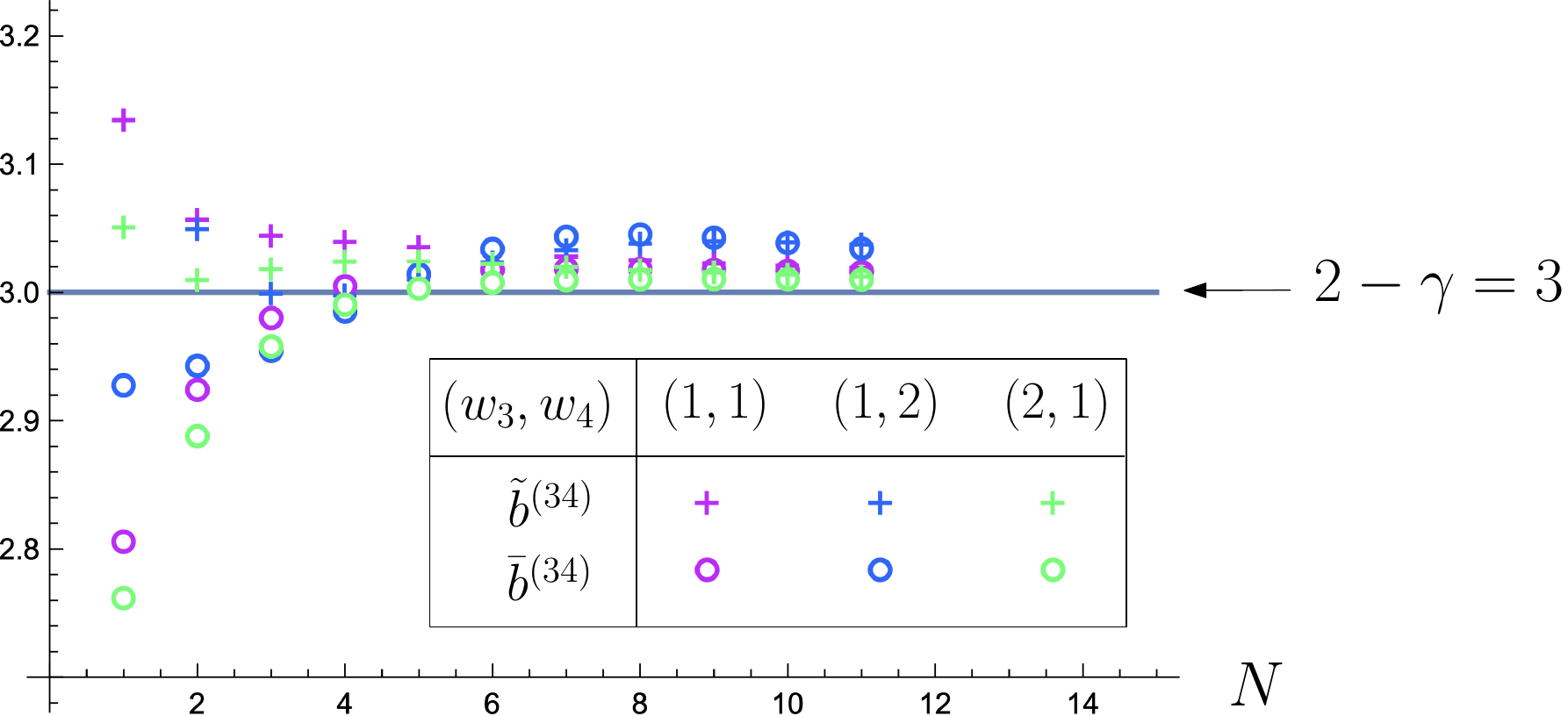}
   \caption{\small Estimates of $2-\gamma$ for Hamiltonian cycles on bicolored planar maps with mixed valencies
   in $\mathcal{S}=\{3,4\}$  with $(w_3,w_4)=(1,1)$, $(1,2)$ and $(2,1)$ respectively.}
  \label{fig:gamma34est}
\end{figure}

Figure~\ref{fig:gamma23-24est} presents our estimates of $2-\gamma$ for bicolored planar maps with mixed valencies for 
$\mathcal{S}=\{2,3\}$ and $\{2,4\}$ (with all weights $w_i=1$) while Figure~\ref{fig:gamma34est} presents our estimates
for bicolored planar maps with mixed valencies in $\mathcal{S}=\{3,4\}$ with $(w_3,w_4)=(1,1),(1,2)$ and $(2,1)$ respectively.
The estimates now agree with the expected value $\gamma=-1$ of Corollary~\ref{claim6}.

\section{Rigid Hamiltonian cycles on $\boldsymbol{2q}$-regular bicolored planar maps}
\label{sec:rigid}
\begin{figure}
  \centering
  \fig{.6}{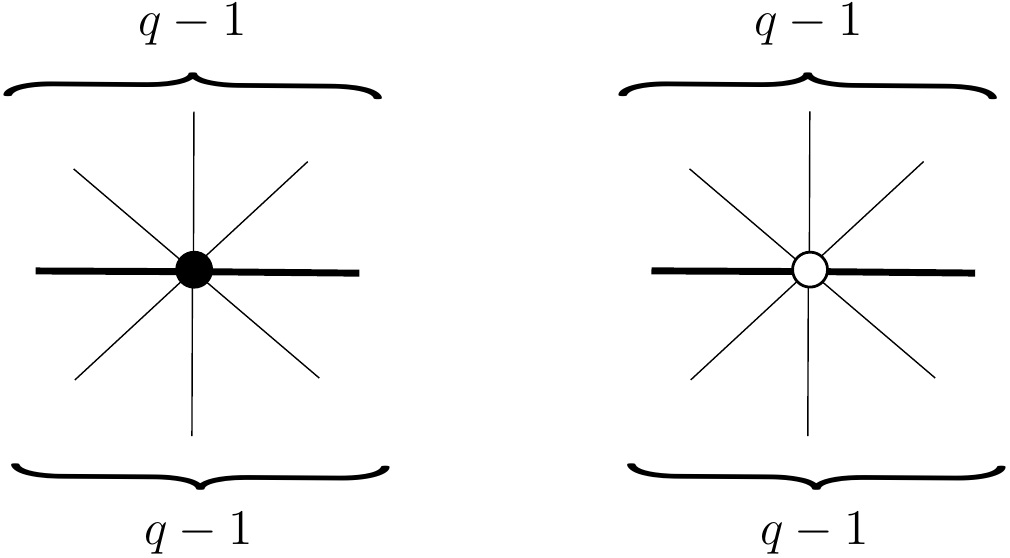}
   \caption{\small Example of the edge environment of a black and of a white vertex in the RFPL model. Each vertex is traversed by a loop (thick edges)
   in such a way that there are exactly $q-1$ unvisited edges (thin edges) on each side of the loop (here $q=4$).}
  \label{fig:rigid}
\end{figure}
\subsection{Definition and properties}
Let us now discuss a restricted class of Hamiltonian cycles, or more generally of fully packed loops, which, as in \cite{BBG11},  we call \emph{rigid}. Those
are defined as follows: a rigid fully packed loop (RFPL) configuration is a set of fully packed loops on a $2q$-regular bicolored planar
map, with $q\geq 2$ a fixed integer, such that, at each vertex, the unvisited edges are equally distributed on both sides of the loop, i.e., with
exactly $(q-1)$ of them on each side, see Figure~\ref{fig:rigid}. As before, each loop receives a weight $n$: this defines the RFPL$(n)$ model
on $2q$-regular bicolored planar maps. Again the $n\to 0$ limit selects configurations of \emph{rigid Hamiltonian cycles}, i.e., 
configurations with a single self-avoiding loop visiting all the vertices of the map. 

For $2q=4$, a rigid Hamiltonian cycle configuration is what was called a \emph{meandric system} in \cite{FT22,BGP22}. Note that 
a $4$-regular planar map equipped with a rigid Hamiltonian cycle is automatically bicolorable. 

Let us again start with the RFPL$(2)$ model, corresponding to (unweighted) oriented loops. As we did in Section~\ref{sec:pregular},
we may distinguish $A$- (unvisited), $B$- (visited oriented towards a black vertex) and $C$- (visited oriented towards a black vertex)
edges, which allows us to assign a $d$-component height $\X\in\R^d$ to each face of the map, whose variation $\Delta\X$ between adjacent 
faces depends on the nature of the edge between them according to the rules of Figure~\ref{fig:heights}.
As before, this height is well-defined by requiring the necessary and sufficient condition (corresponding to \eqref{eq:sumrule} for $p=2q$):
\begin{equation}
2(q-1)\A+\B+\C=\boldsymbol{0}
\label{eq:sumrule2q}
\end{equation}
which, de facto, fixes $d=2$, with $\X$ living in the $(\B,\C)$-plane with $\B$ and $\C$ two unit vectors with, say $\B\cdot\C=-1/2$.
As before, it is convenient to express $\X$ in the orthogonal basis $(\A,\btwo)$, with $\btwo:=\B-\C$ and, as in Section~\ref{sec:pregular}, 
write the associated coarse grained average value $\boldsymbol{\Psi}  = \langle \X \rangle$ as a
two-component vector field $\boldsymbol{\Psi}=\psi_1\A+\psi_2 \btwo$ with components both along $\A$ and along $\btwo$.
Reproducing the arguments of Section~\ref{sec:pregular}, it would be tempting to infer that the results of Claims~\ref{claim1}
and \ref{claim2} hold, i.e., that the RFPL$(n)$ model is the coupling to gravity of a CFT of central charge $c_{\mathrm{fpl}}(n)$.
We will now argue that this conclusion is actually incorrect and that the RFPL$(n)$ model is the coupling to gravity of a CFT of central charge 
$c_{\mathrm{dense}}(n)=c_{\mathrm{fpl}}(n)-1$. Indeed, even though we may define the coordinate $\psi_1$ in the $\A$ direction, the value
of this coordinate is in practice \emph{frozen}, equal to a fixed value (which we may take equal to $0$) on the entire map.
We thus state:
\begin{trueprop}
\label{claim7}
The two-component vector field $\boldsymbol{\Psi}$ varies only via its coordinate $\psi_2$ along the $\btwo$ direction,
which makes it in practice a one-component vector field.
\end{trueprop}
This de facto reduces the central charge by $1$, hence we arrive at:
\begin{prop}
\label{claim8}
For $-2\leq n\leq 2$, the RFPL$(n)$ model on $2q$-regular bicolored planar maps is described by the coupling to gravity of a CFT
with central charge 
\begin{equation}
c=c_{\mathrm{dense}}(n)=1-6\frac{(1-g)^2}{g}\quad \mathrm{where}\  n=-2\cos(\pi\, g)\ \mathrm{with}\ 0\leq g\leq 1 \ .
\end{equation}
\end{prop} 
In the $n\to0$ limit, we deduce in particular:
\begin{cor}
\label{claim9}
The model of rigid Hamiltonian cycles on $2q$-regular bicolored planar maps is described by the coupling to gravity of a CFT
with central charge 
\begin{equation}
c=c_{\mathrm{dense}}(0)=-2\ .
\end{equation}
In particular, using KPZ \eqref{eq:KPZ}, the associated partition function $z_N$ has the 
asymptotic behavior \eqref{eq:asympz} with
\begin{equation}
\gamma=\gamma(-2)=-1.
\end{equation}
\end{cor} 

\subsection{Proof of Proposition~\ref{claim7}}
\begin{figure}[htbp]
  \centering
  \fig{.9}{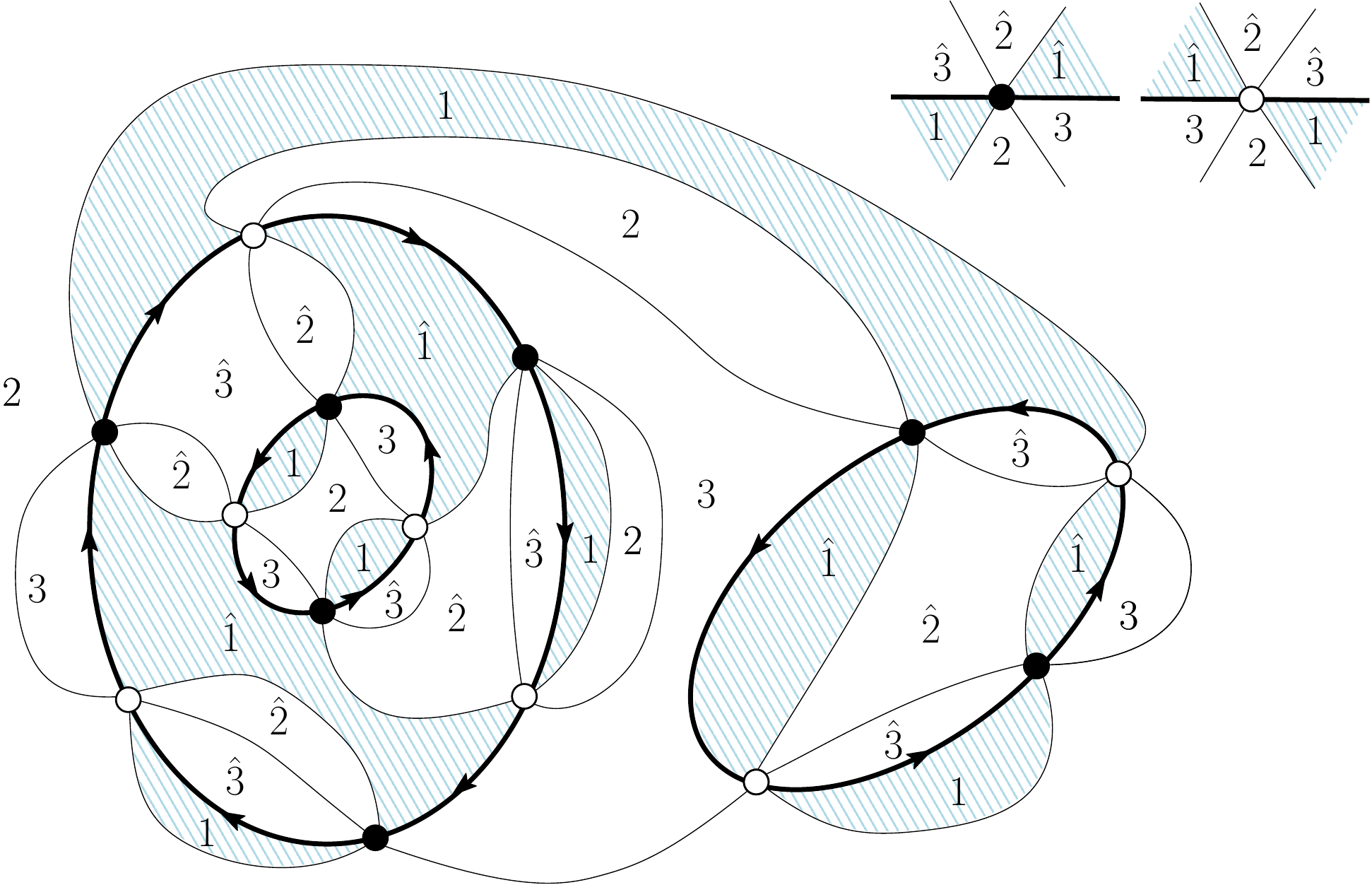}
   \caption{\small The splitting of the face set into subsets $\mathcal{F}_1,\mathcal{F}_2,\mathcal{F}_3,\mathcal{F}_{\hat{1}},
\mathcal{F}_{\hat{2}},\mathcal{F}_{\hat{3}}$ for a $6$-regular bicolored planar map. The order of appearance of the faces is 
$(1,2,3,\hat{1},\hat{2},\hat{3})$ 
clockwise around white vertices and counterclockwise around black vertices (as shown the upper right corner) and, in the presence of \emph{rigid} fully packed loops,
we may chose the numbering so that the loops always separate faces labelled $1$ from faces labelled $\hat{3}$ 
and faces labelled $\hat{1}$ from faces labelled $3$. }
  \label{fig:p6rigid}
\end{figure}
\begin{proof}The following argument is a generalization to arbitrary $q$ of that given in \cite[Sect.~11.3]{DFG05} for the case $q=2$.
The first remark is that the set of faces of a bicolored $p$-regular planar map is naturally split into $p$ subsets as follows\footnote{The reader might be
more familiar with the dual picture: bicolored $p$-regular planar maps are dual to planar Eulerian $p$-angulations (with 
bicolored black and white faces all of valency $p$), a particular instance of $p$-constellations \cite{BMS00}.}:
pick a reference face $f_0$ and label each face $f$ of the map by $\ell(f) = (L(f) \mod p) +1$ where $L(f)$ is the number of crossed edges of any path
connecting $f_0$ to $f$ and \emph{traversing only edges with their white vertex on the right} (or equivalently turning clockwise around white vertices and
counterclockwise around black ones). It is easily seen that $L(f)$ is indeed independent on the chosen path. This splits the set of faces into 
$p$-subsets which we denote by $\mathcal{F}_1,\mathcal{F}_2,\ldots,\mathcal{F}_p$ where $\mathcal{F}_j$ is the set of faces labelled $j$. 
Moreover, it is easily seen that, by construction, the cyclic order of the labels is $(1,2,\ldots, p)$ both clockwise around white vertices and counterclockwise 
around black ones. For $p=2q$, we may 
instead use labels $\ell \in \{1,2,\ldots,q,\hat{1},\hat{2},\ldots, \hat{q}\}$ so that the subsets are now denoted by $\mathcal{F}_1,\mathcal{F}_2,\ldots,\mathcal{F}_q,\mathcal{F}_{\hat{1}},
\mathcal{F}_{\hat{2}},\ldots,\mathcal{F}_{\hat{q}}$ and the cyclic order of the labels is $(1,2,\ldots, p,\hat{1},\hat{2},\ldots, \hat{q})$. In the presence of
rigid fully packed oriented loops, we may finally choose the face $f_0$ so that the loops always separate faces in $\mathcal{F}_1$  from faces in $\mathcal{F}_{\hat{q}}$ 
and faces in $\mathcal{F}_{\hat{1}}$ from faces in $\mathcal{F}_q$ (it is enough to impose this property at one vertex and, since the loops are rigid, it
automatically propagates\footnote{Note that the set $\mathcal{F}_1\cup\mathcal{F}_{\hat{1}}$ needs not be connected. Still, one can check that the property
propagates from one connected component to the other. This is because the edges separating $\mathcal{F}_j$ from $\mathcal{F}_{j-1}$ and $\mathcal{F}_{\hat{j}}$ from $\mathcal{F}_{\widehat{j-1}}$ for
any given $j\in\{2,\ldots q\}$ also form a set of rigid fully packed loops.} to all the vertices), see Figure~\ref{fig:p6rigid} for an example in the case $q=3$.
 
\begin{figure}
  \centering
  \fig{.7}{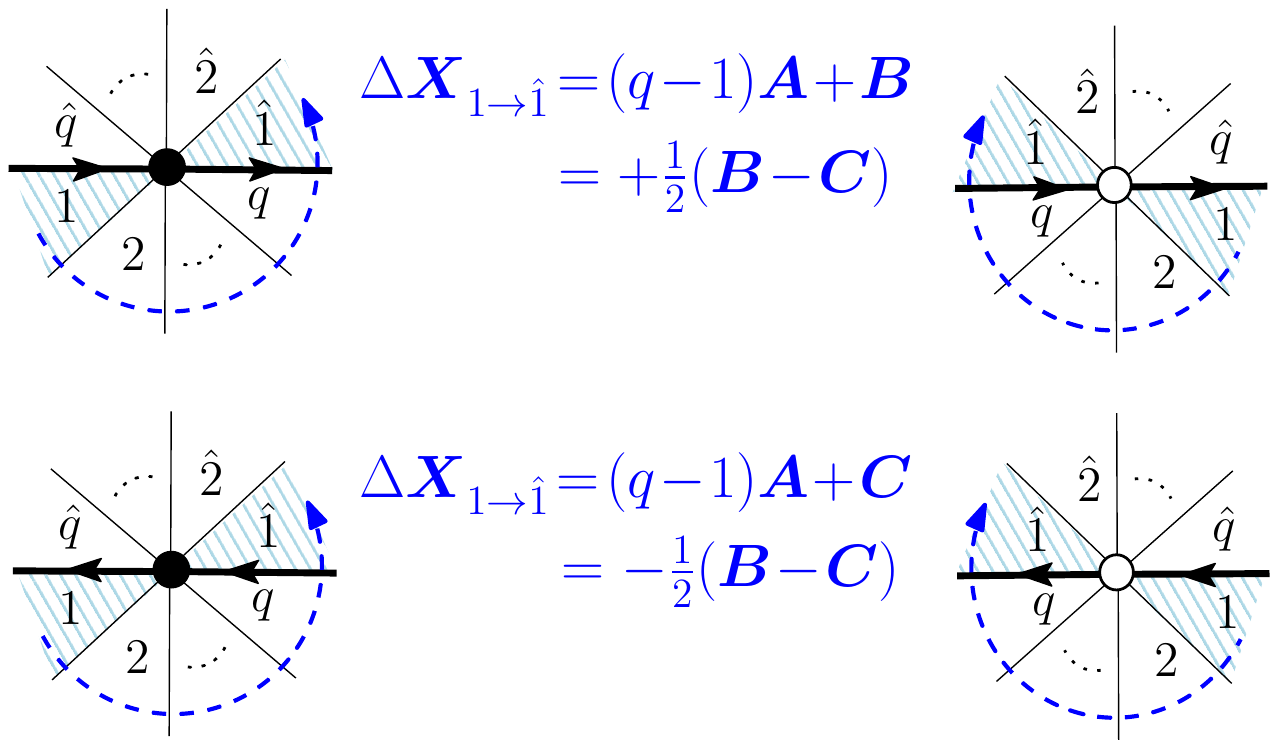}
   \caption{\small The change of height $\Delta X_{1\to \hat{1}}$ at a given vertex when going from the face with label $1$ to that, opposite, with label $\hat{1}$
is given by $\Delta X_{1\to \hat{1}}=(q-1)\A+\B$ or $\Delta X_{1\to \hat{1}}=(q-1)\A+\C$ depending on the orientation of the loop. From the relation \eqref{eq:sumrule2q},
$\Delta X_{1\to \hat{1}}$ is therefore equal to $\pm \frac{1}{2} (\B-\C)=\pm \frac{1}{2} \btwo$, with no component along the $\A$ direction.}
  \label{fig:rigiddeltaX}
\end{figure}
Focusing now on the subset 
$\mathcal{F}_1\cup\mathcal{F}_{\hat{1}}$, we observe that, as shown in Figure~\ref{fig:rigiddeltaX}, the change of height $\Delta X_{1\to \hat{1}}$  when going from of the face with label $1$ to that with label $\hat{1}$ at a given vertex
is always given by
\begin{equation}
\Delta X_{1\to \hat{1}}=\pm \frac{1}{2} (\B-\C) =\pm \frac{1}{2} \btwo\ ,
\end{equation}
with a sign depending on the orientation of the loop.
Note finally that, \emph{at any given vertex}, the change of height $\Delta X_{\ell\to \hat{\ell}}$  when going from the incident face with label $\ell$ to that with label $\hat{\ell}$ is in practice
\emph{independent of $\ell$}: all in all, the coarse grained height $\boldsymbol{\Psi}$ (whatever its precise definition) has only variations in the $\btwo$ direction.
\end{proof}
\subsection{Exact enumeration}
\label{eq:exactenum}
\begin{figure}
  \centering
  \fig{.9}{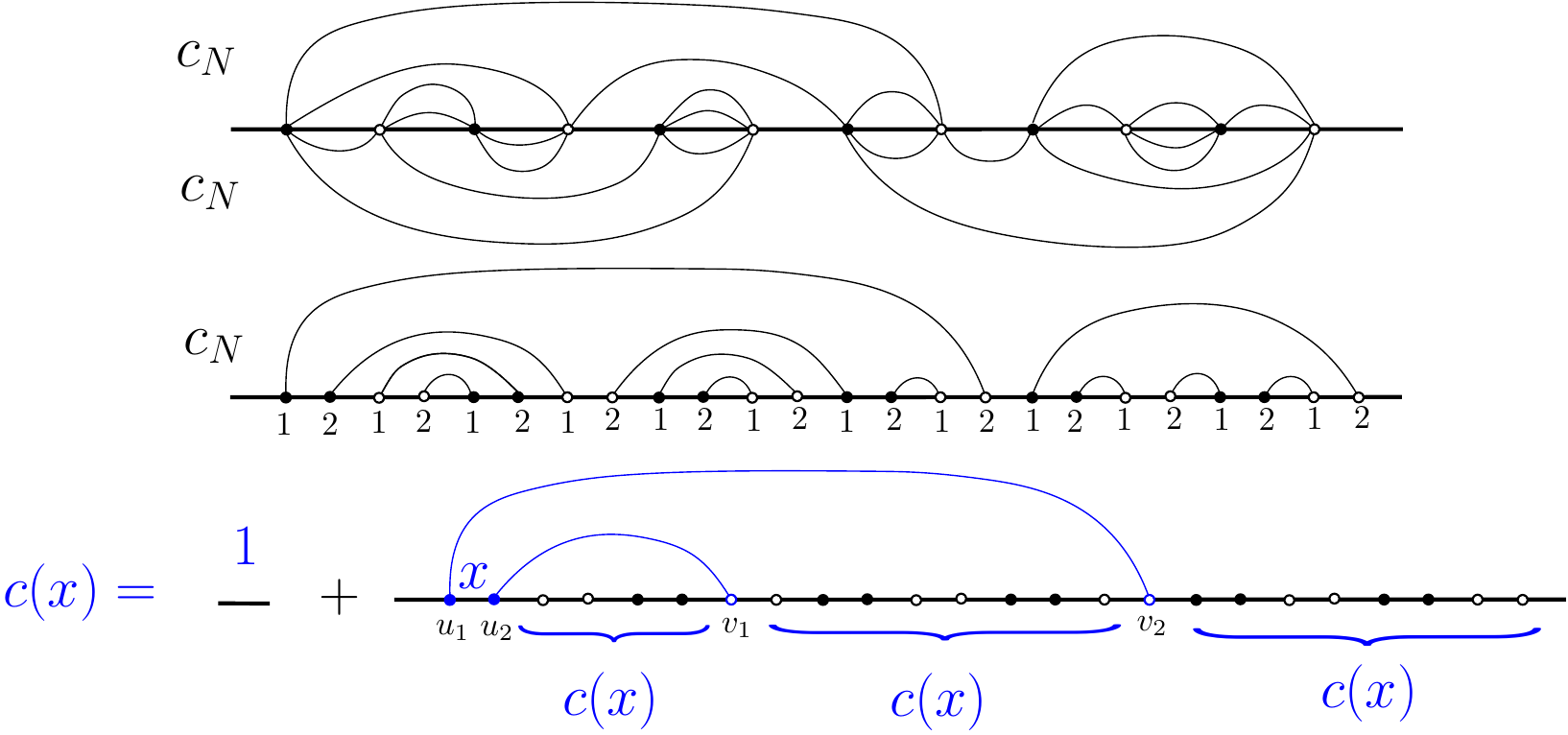}
   \caption{\small Top: an example of rigid Hamiltonian cycle on a $2q$-valent bicolored planar map (here with $q=3$), after opening it into
   a straight line of alternating black and white vertices. The upper and lower parts are independent arch systems, both enumerated
   by $c_N$. Middle: alternative representation of the upper arch system after splitting each vertex into $(q-1)$ successive copies of the same color.
   A black (resp. white) vertex labelled $j$ is connected to a white (resp. black) one labelled $q-j$ (here with $q=3$ and $j=1,2$). Bottom: schematic picture 
    of the decomposition of an arch configuration enumerated by $c(x)$ (with a weight $x$ per group of $q$ arches) into $q$ sequences 
    of arch configurations, each of them also enumerated by $c(x)$. Note that the order of colors within different subsequences is always the same, up
    to a cyclic permutation.}
  \label{fig:fusscatalan}
\end{figure}

The fact that $\gamma=\gamma(-2)=-1$ for rigid Hamiltonian cycles on $2q$-regular bicolored planar maps may be checked by an exact enumeration
of the allowed configurations. By embedding the map on the Riemann sphere, i.e., opening the cycle into a straight line of alternating black and white vertices, we immediately see that the rigidity
constraints (imposing that the number of unvisited edges incident to any vertex is $(q-1)$ \emph{on each side} of the straight line) allows us to write by symmetry
\begin{equation}\label{eq:zc2}
z_N=c_N^2 \ ,
\end{equation}
where $c_N$ enumerates configurations of non-crossing bicolored arches connecting the black and white vertices on one side of the straight line only, each vertex
being incident to exactly $(q-1)$ arches, see Figure~\ref{fig:fusscatalan}-top for an illustration. 

As for $c_N$, it is easily evaluated from the following argument: start by splitting
each vertex into $(q-1)$ copies of the same color, with one arch incident to each copy, the choice of the arch to be connected being entirely dictated by the non-crossing constraint of the arches. 
We now have a sequence made of $N$ groups of $(q-1)$ successive black vertices alternating with $N$ groups of $(q-1)$ successive white vertices. 
In a given monocolor group of size $(q-1)$, we may label the vertices from $1$ to $(q-1)$ from left to right: the non-crossing constraint imposes that a black vertex with label $j$ 
is necessarily connected to a white vertex with label $(q-j)$ for any $j\in\{1,\ldots,q-1\}$, see Figure~\ref{fig:fusscatalan}-middle.

Looking now at the $(q-1)$ first black
vertices on the left, denoted by $u_1,\ldots,u_{q-1}$ (so that $u_j$ has the abovementioned label $j$) and calling $v_{q-j}$ the white vertex to which $u_j$ is connected (so that $v_{q-j}$ has the 
abovementioned label $(q-j)$), these latter vertices split the remaining $2(q-1)(N-1)$ vertices into subsequences respectively between $u_{q-1}$ and $v_{1}$, between $v_{1}$ and $v_{2}$, $\ldots$,
between $v_{q-2}$ and $v_{q-1}$, and finally to the right of $v_{q-1}$. This yields a total of $q$ subsequences of non negative integer lengths $2(q-1)m_1,\ldots,2(q-1)m_q$ respectively with $m_j\geq 0$
for $j=1,\ldots,q$. Due to the presence of the $(q-1)$ first arches, each of these $q$ subsequences is separated from the others: in particular, the pairing by arches of the vertices 
takes place independently within each subsequence. 
Moreover, \emph{at the price of a cyclic permutation} of its vertices, the $j$-th subsequence is made of $m_{j}$ groups of $(q-1)$ successive 
black vertices alternating with $m_{j}$ groups of $(q-1)$ successive white vertices, see Figure~\ref{fig:fusscatalan}-bottom. The number of possible arch configurations for the $j$-th subsequence is therefore given by $c_{m_{j}}$ (independently of the required cyclic permutation).
We arrive at the recursion relation
\begin{equation}
c_N=\sum_{m_1,\ldots,m_q\geq 0 \atop m_1+\cdots+m_q=(N-1)} \prod_{j=1}^{q} c_{m_{j}}\ , \quad N\geq 1
\end{equation}  
with the convention $c_0=1$. Introducing the generating function $c(x):=\sum_{N\geq 0}c_N\, x^N$, we deduce that
\begin{equation}
c(x)=1+x\left(c(x)\right)^q\, ,
\end{equation}
where we recognize the equation determining the generating function $c(x)$ of the $q$-th generalized Fuss-Catalan numbers \cite{Fuss}
\begin{equation}
c_N=\frac{1}{(q-1)N+1}{q\, N\choose N}\ , \quad N\geq 0\ .
\label{eq:fusscatalan}
\end{equation}
In particular, when $q=2$, we recover the celebrated Catalan numbers.
As a consequence of \eqref{eq:fusscatalan}, we get
\begin{equation}
z_N=\left(\frac{1}{(q-1)N+1}{q\, N\choose N}\right)^2\underset{N\to \infty}{\sim} \frac{q}{2\pi (q-1)^3} \frac{ \left(\frac{q^q}{(q-1)^{q-1}}\right)^{2N} } { N^{3} }\ .
\label{eq:asymprigid}
\end{equation}
As expected, $z_N$ has the asymptotic behavior \eqref{eq:asympz}
with 
\begin{equation}
\gamma=\gamma(-2)=-1\ , \quad \mu=\frac{q^q}{(q-1)^{q-1}}\quad \mathrm{and}\ \varkappa=  \frac{q}{2\pi (q-1)^3}\ .
\end{equation}

\section{Long-distance contacts within Hamiltonian cycles}
\label{sec:middle}
\subsection{Scaling limits of the $\mathrm{O}\boldsymbol{(n)}$ and $\mathrm{FPL}\boldsymbol{(n)}$ models on regular lattices} \label{sec:scalinglimit}
It is widely believed that the \emph{scaling limit} of the critical $\mathrm{O}(n)$ model on two dimensional regular (e.g., hexagonal or square) lattices is described by the celebrated \emph{Schramm-Loewner evolution} $\mathrm{SLE}_\kappa$ \cite{OS,MR2153402}, 
and its collection of critical loops by the so-called \emph{conformal loop ensemble} $\mathrm{CLE}_\kappa$ \cite{SS09}. This conformally invariant random process depends on a single parameter $\kappa \geq 0$, 
which in the $\mathrm{O}(n)$ model case is  $\kappa=4/g$ \cite{SS09,Duplantier03,MR2112128,zbMATH05541623} so that :
\begin{equation}\label{eq:nkappa}
\begin{split}
n=-2\cos (4\pi /\kappa),\,\,\,&\kappa\in [8/3,4] \quad \hbox{for the \emph{dilute} critical point},\\
&\kappa\in (4,8]\,\,\,\,\,\quad  \hbox{for the \emph{dense} critical phase}.
\end{split}
\end{equation}  
For $n\in (0,2]$, one has $\kappa\in (8/3, 8)$, i.e., the range for which $\mathrm{CLE}_\kappa$ is defined, whereas the SLE$_\kappa$ process is actually defined for $\kappa\in [0,\infty)$. Note that for
$n\to 0$,  in the dilute case, the limit of $\mathrm{CLE}_\kappa$ as $\kappa \searrow 8/3$ is $\mathrm{SLE}_{8/3}$ and, in the dense case, the limit of $\mathrm{CLE}_\kappa$ as $\kappa \nearrow 8$ is space-filling $\mathrm{SLE}_{8}$.
The full critical  $\mathrm{O}(n)$ model  range $n\in[-2,2]$ corresponds to $\kappa\in [2,\infty)$
$\mathrm{SLE}_{\kappa}$ paths, which are always non self-crossing, are \emph{simple}, i.e., non-intersecting when $\kappa \in [0,4]$, and \emph{non-simple}  
when $\kappa \in (4,\infty)$ \cite{MR2153402}. 

This scaling limit has been rigorously established in several  cases: the uniform spanning tree for which $n=0, g=1/2, \kappa=8$ \cite{LSW03}; the loop-erased random walk for which (formally) $n=-2, g=2, \kappa=2$ \cite{LSW03,LV21}; the contour lines of the discrete Gaussian free field, for which $n=2, g=1, \kappa=4$ \cite{10.1007/s11511-009-0034-y}; critical site percolation on the triangular  lattice \cite{SMIRNOV2001239,Camia2006}, for which $n=1, g=2/3, \kappa=6$; the critical Ising model and its associated Fortuin-Kasteleyn random cluster model on the square lattice \cite{zbMATH05808591,Chelkak2012515} for which,   respectively, $n=1, g=4/3, \kappa =3$ and $n=\sqrt{2}, g=3/4, \kappa=16/3$. 

The associated $\mathrm{SLE}_\kappa$ central charge is then 
\begin{equation}\label{eq:ckappa}
c=c_{\mathrm{sle}}(\kappa):=\frac{1}{4}(6-\kappa)\left(6-\frac{16}{\kappa}\right) \in (-\infty, 1]\quad \mathrm{for} \quad \kappa >0 \ .
\end{equation}
Notice the invariance of the central charge \eqref{eq:ckappa} under the SLE$_\kappa$  \emph{duality} \cite{Duplantier00,Duplantier03,MR2112128,MR2439609,dub_dual},
\begin{equation}\label{eq:kduality}
\kappa \leftrightarrow 16/\kappa=: \widetilde \kappa \ .
\end{equation} 
 The geometrical interpretation of this  duality is as follows. 
 In the scaling limit, loops in the dense $\mathrm{O}(n)$ model are \emph{non-simple} paths of Hausdorff dimension \cite{PhysRevLett.49.1062,zbMATH03959008}  
 $D=1+(2g)^{-1}=1+\kappa/8> 3/2$ for $g\in [1/2,1),\,\,\kappa\in (4,8]$ ; their \emph{external perimeters} are \emph{simple} critical paths of Hausdorff dimension  \cite{Duplantier00} $\widetilde D=1+g/2=1+\widetilde \kappa/8< 3/2$. These Hausdorff dimensions thus satisfy the 
  universal duality relation 
  \begin{equation}\label{eq:duality}
  (D-1)(\widetilde D-1)=\frac{1}{4}\ ,
  \end{equation} which has been directly established for critical percolation \cite{PhysRevLett.83.1359}. Non-simple $\mathrm{SLE}_{\kappa}$ paths   
for $\kappa \in (4,8]$ have indeed been proven to have for outer 
boundaries  dual simple SLE$_{\widetilde\kappa}$ paths, with $\widetilde \kappa=16/\kappa\in [2,4)$ \cite{MR2439609,dub_dual}.

The so-called \emph{watermelon} exponents  (conformal weights) corresponding to the merging of a number $\ell$ of  conformally invariant SLE$_\kappa$ paths \cite{MR2112128}, in particular of $\ell$ critical lines in the (dense or dilute) $\mathrm{O}(n)$ model with $n$ as in \eqref{eq:nkappa} are given by \cite{PhysRevLett.49.1062,zbMATH03959008,DG1987,0305-4470-19-13-009,0305-4470-19-16-011,Duplantier1987291,dupJSP97,PhysRevLett.58.2325,PhysRevLett.61.138}
\begin{equation}\label{eq:hkell}
h_\ell^{(\kappa)}=\frac{1}{16\kappa}\left[4\ell^2-(4-\kappa)^2\right], \quad \ell \in \mathbb Z^+.
\end{equation} 
As anticipated above, the Hausdorff dimension of $\mathrm{SLE}_{\kappa}$ is  \cite{AOP364} 
\begin{equation}\label{eq:D}
D=\inf\{2(1-h_2^{(\kappa)}),2\}=\inf\{1+\kappa/8,2\}\ .
\end{equation}

The fully-packed $\mathrm{FPL}(n)$ model on the hexagonal lattice \cite{BN94,BSY94,KdGN96} or on the square lattice \cite{BBNY96,JK98} is related to the corresponding dense $\mathrm{O}(n)$ model via a shift of its central charge by one unit as in 
\eqref{eq:valcgn} and \eqref{eq:valcgndense}. 
The watermelon exponents for an \emph{even} number of paths are the same in  $\mathrm{FPL}(n)$ and dense $\mathrm{O}(n)$ models, and in particular the 2-leg exponent which gives the Hausdorff dimension of the paths, whereas those for a \emph{odd} number of paths differ both on the hexagonal ({\Large$\hexagon$}) \cite{BN94,BSY94,KdGN96}, and on the square ($\square$) \cite{BBNY96,JK98} lattices,
\begin{equation}
  \begin{split}
&h^{\mathrm{fpl}(n)}_{2k}=h_{2k}^{(\kappa)},\\
&h^{\mathrm{fpl}(n)}_{2k-1}=h_{2k-1}^{(\kappa)}+\frac{3}{4\kappa} \ \ \ ({\Large\hexagon}),\\
&h^{\mathrm{fpl}(n)}_{2k-1}=h_{2k-1}^{(\kappa)}+\frac{1}{6+\kappa} \ \ \ (\square),\quad k\in \mathbb Z^+.
\end{split}
\label{eq:discry}
\end{equation}  
Even in the presence of the mismatch of central charges \eqref{eq:valcgn} and \eqref{eq:valcgndense}, one is thus led to conjecture \cite{D2G2,BGS22,BGP22} that the scaling limit of the fully-packed $\mathrm{FPL}(n)$ loop model itself on the honeycomb or square lattices is described by a conformal loop ensemble $\mathrm{CLE}_\kappa$, with  $\kappa$ corresponding to the dense $\mathrm{O}(n)$ model phase \cite{Re91,BN94,BSY94,KdGN96,BBNY96,JK98},
\begin{equation}\label{eq:kappan}
\kappa=\frac{4\pi}{\arccos(-n/2)} \in (4,8]\quad \mathrm{for} \quad n\in [0,2)\ .
\end{equation}
\subsection{Scaling limit for Hamiltonian cycles}
\begin{figure}
  \centering
  \fig{.6}{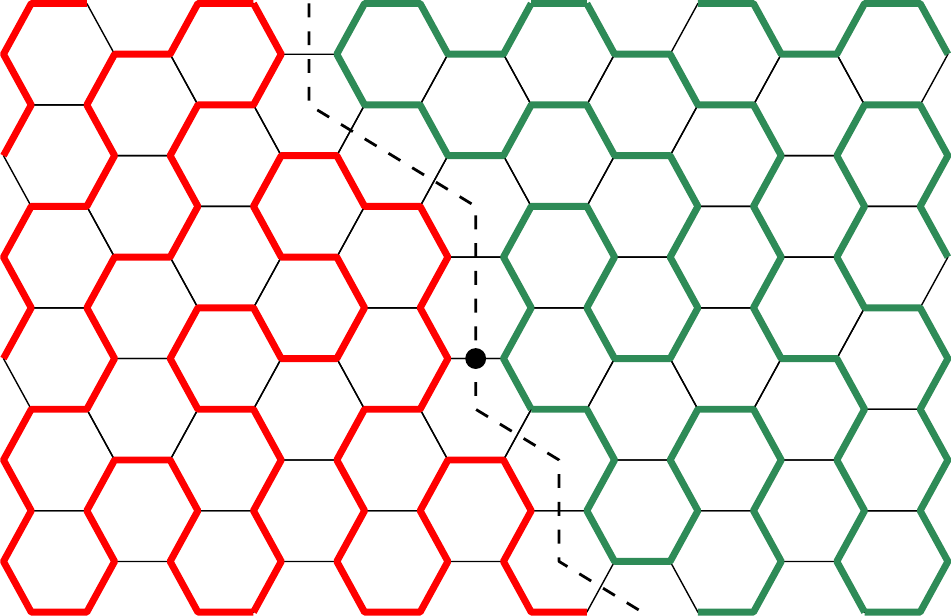}
   \caption{\small  On the hexagonal lattice with the spherical topology, the two (red and green) \emph{halves} $\mathcal C_1$ and 
   $\mathcal C_2$ of a \emph{Hamiltonian cycle} $\mathcal C=\mathcal C_1 \cup \mathcal C_2$ are separated by a (dotted) dual  
   loop $\widetilde {\mathcal C}=\mathcal C_1 \cap \mathcal C_2$ on the dual lattice that crosses the whole set of their \emph{contact links}.  
   This separatrix can be seen as the \emph{external perimeter} of each half of $\mathcal C$. A point along that dual loop can be viewed as the origin of either $\ell=4$ compact $\mathrm{O}(n=0)$ half-lines, or 
   of $\ell=2$ dual half-lines. In the scaling limit, the fully-packed loop $\mathcal C$ converges to space-filling SLE$_{\kappa=8}$ with Hausdorff dimension $D=2$, and its fractal contact set $\widetilde{\mathcal C}$ to whole-plane 
   SLE$_{\widetilde\kappa=2}$, with Hausdorff dimension $\widetilde D=5/4$.}
  \label{fig:separatrice}
\end{figure}
Let us now consider the $\mathrm{FPL}(n=0)$ case of a single Hamiltonian cycle $\mathcal C$ with $2N$ vertices, drawn on the regular bicolored hexagonal (or square) lattice, with the sphere topology. Marking two points at distance $N$ along $\mathcal C$ splits this cycle into two equal parts $\mathcal C_i, i=1,2$  such that $\mathcal C=\mathcal C_1\cup \mathcal C_2$. They are separated by a single closed path $\widetilde{\mathcal C}$ drawn on the dual triangular lattice, that crosses the whole set of \emph{contacts links}, i.e., edges incident to a vertex in $\mathcal C_1$ and to one in $\mathcal C_2$. We write 
$\widetilde{\mathcal C}=\mathcal C_1 \cap \mathcal C_2$ by a slight abuse of notation.  In the spherical topology, this dual path can be viewed as the common external perimeter shared by each of the two halves $\mathcal C_i, i=1,2$ of  $\mathcal C$  (see Figure \ref{fig:separatrice}).

In the scaling limit, one has $g=1/2, \kappa=8$, so the cycle $\mathcal C$  should converge to a \emph{conformally invariant} $\mathrm{SLE}_8$ path drawn on the Riemann sphere, which is a \emph{Peano curve}, i.e.,  a space-filling curve with Hausdorff dimension $D=2$.  By duality \eqref{eq:kduality} \eqref{eq:duality}, the path $\widetilde{\mathcal C}$ should then converge to a whole-plane SLE$_2$ curve with Hausdorff dimension $\widetilde D=5/4$. 

This can be directly checked by observing that a contact point on $\widetilde{\mathcal C}$ can be viewed as the origin of  $\ell=4$ fully-packed $n=0$ lines, i.e., in the scaling limit, that of $\ell=4$ space-filling $\mathrm{SLE}_8$ paths, as well as the origin of $\ell=2$ 
$\mathrm{SLE}_2$ dual paths, with identical conformal weights  \eqref{eq:hkell}
\begin{equation}\label{eq:hhdual}
h_{1\cap2}:=h^{\mathrm{fpl}(0)}_{\ell=4}=h_{\ell=4}^{(\kappa=8)}=h_{\ell=2}^{(\widetilde{\kappa}=2)}= \frac{3}{8}\ . 
\end{equation}
The expected number $|\widetilde{\mathcal C}|=|\mathcal C_1 \cap \mathcal C_2|$ of contact links between the two halves of Hamiltonian cycle $\mathcal C$,  in a large domain $\mathcal D$ of area $A=|\mathcal D|$ on the regular bicolored lattice, is then given, in the scaling limit, by
\begin{equation}
\label{eq:contactexp}
\mathbb E\, |\mathcal C_1 \cap \mathcal C_2| \asymp A^{\widetilde{D}/2} = A^{1-h_{1\cap2}},\,\,\,h_{1\cap2}=3/8, \quad A\to \infty\ ,
\end{equation}
where the asymptotic equivalence $\asymp$ means that the ratio of logarithms tends to 1. 
\subsection{Coupling to quantum gravity}\label{sec:KPZ} Random planar maps, as weighted by the partition functions of critical statistical models, are widely believed to have for scaling limits \emph{Liouville quantum gravity} (LQG) coupled to the conformal field theory describing these critical models \cite{KPZ88,FD88,DK89}, or, equivalently, to the corresponding SLE processes \cite{10.1214/15-AOP1055,10.1214/15-AOP1061,PhysRevLett.107.131305,DMS21}.
The continuum description of the random planar map area involves a (regularized) Liouville quantum measure $d^2x\, :e^{\gamma_{\scriptscriptstyle{\mathrm{L}}} \varphi_{\scriptscriptstyle{\mathrm{L}}}(x)}:$ 
in terms of a Gaussian free field (GFF) $\varphi_{\scriptscriptstyle{\mathrm{L}}}$ \cite{springerlink:10.1007/s00222-010-0308-1bis}, possibly weighted as in the Liouville 
action \cite{KPZ88,FD88,DK89}. For the coupling to gravity of a CFT with central charge $c$, the Liouville parameter $\gamma_{\scriptscriptstyle{\mathrm{L}}}$ is \cite{KPZ88,FD88,DK89,DMS21,10.1214/15-AOP1055,10.1214/15-AOP1061,PhysRevLett.107.131305}
 \begin{equation}
\gamma_{\scriptscriptstyle{\mathrm{L}}}=\gamma_{\scriptscriptstyle{\mathrm{L}}}(c):=\frac{1}{\sqrt{6}}\left(\sqrt{25-c}-\sqrt{1-c}\right)\in (0,2] \quad \mathrm{for}\quad c\in (-\infty, 1]\ .
\label{eq:gammaL}
\end{equation}
An Euclidean fractal measure associated with a set of Hausdorff dimension $D=2(1-h)$ is transformed in LQG into a quantum fractal measure, via a local multiplicative factor 
of the form $:e^{\alpha\varphi_{\scriptscriptstyle{\mathrm{L}}}}:$ with $\alpha:=\gamma_{\scriptscriptstyle{\mathrm{L}}}(1-\Delta)$, where the quantum scaling exponent $\Delta$ is the analogue of the Euclidean scaling exponent $h$ \cite{FD88,DK89,PhysRevLett.107.131305}. It is given 
by the celebrated  KPZ relation \cite{KPZ88}, 
\begin{equation}
\Delta=\Delta(h,c):=\frac{\sqrt{1-c+24\, h}-\sqrt{1-c}}{\sqrt{25-c}-\sqrt{1-c}}\ ,
\label{eq:Deltahc}
\end{equation}
in terms of the original scaling exponent $h$  (e.g., conformal weight) of the CFT of central charge $c$. 
Eq. \eqref{eq:Deltahc} can be inverted with the help of the Liouville parameter \eqref{eq:gammaL} as the simple quadratic formula,
\begin{equation} \label{eq:KPZdirect}
h(\Delta)=\frac{\gamma_{\scriptscriptstyle{\mathrm{L}}}^2}{4}\Delta^2+\left(1-\frac{\gamma_{\scriptscriptstyle{\mathrm{L}}}^2}{4}\right)\Delta\ .
\end{equation}
Its rigorous proof \cite{springerlink:10.1007/s00222-010-0308-1bis,2009arXiv0901.0277D,rhodes-2008,DRSV12} rests on the assumption that the GFF or  Liouville field  $\varphi_{\scriptscriptstyle{\mathrm{L}}}$ and (any)  random fractal curve (possibly described by a CFT) are \emph{independently} sampled. 

The other KPZ result \eqref{eq:KPZ}   for $\gamma(c)$, the configuration or ``string susceptibility exponent'' 
\begin{equation}
\gamma=1-4/\gamma_{\scriptscriptstyle{\mathrm{L}}}^2, 
\end{equation}
or equivalently \eqref{eq:gammaL} for $\gamma_{\scriptscriptstyle{\mathrm{L}}}(c)$, gives the precise coupling between the LQG and CFT or SLE parameters. By substituting the SLE central charge $c=c_{\mathrm{sle}}(\kappa)$ \eqref{eq:ckappa}, one indeed obtains the simple expressions 
\begin{equation} \label{eq:ggLk} 
\gamma=1-\sup\{4/\kappa, \kappa/4\},\quad
\gamma_{\scriptscriptstyle{\mathrm{L}}}=\inf\{\sqrt{\kappa},\sqrt{16/\kappa}\} \ .
\end{equation}
This has been rigorously established in the probabilistic approach by coupling the Gaussian free field in LQG with SLE martingales \cite{10.1214/15-AOP1055,PhysRevLett.107.131305}.  In the scaling limit, random cluster models on random planar maps can then be shown to converge  (in the so-called peanosphere topology of the mating of trees perspective) to LQG-SLE  \cite{DMS21,gwynne2019mating}. 

This \emph{matching} property \eqref{eq:ggLk} of $\gamma$, $\gamma_{\scriptscriptstyle{\mathrm{L}}}$ and $\kappa$ applies to the scaling limit of the critical, dense or dilute, $\mathrm{O}(n)$ model on a random planar map, as well as  
to the fully-packed $\mathrm{FPL}(n)$ model on random (non bicolored) cubic maps \cite{D2G2}. In the case of the fully-packed model on random \emph{bicolored} maps, this also holds in the case of \emph{mixed valencies} (Claim \eqref{claim5}), or in the \emph{rigid} case of $2q$-regular maps (Claim \eqref{claim8}), with
 \begin{equation} \label{eq:cfplck}
 c=c_{\mathrm{dense}}(n)=c_{\mathrm{sle}}(\kappa)\ .
\end{equation}
However, for random \emph{bicubic} planar maps, as seen in Ref. \cite{D2G2}, and for the general \emph{non-rigid} case of \emph{$p$-regular bicolored} planar maps (Claim \eqref{claim2}), the correspondence \eqref{eq:ggLk} no longer holds, and one then has a \emph{mismatch} \cite{BGS22,BGP22}, with  \eqref{eq:cfplck}  replaced  in \eqref{eq:KPZ},  \eqref{eq:gammaL} and \eqref{eq:Deltahc} by 
 \begin{equation}
 \label{eq:cfplck1} c=c_{\mathrm{fpl}}(n)=1+c_{\mathrm{sle}}(\kappa)\ ,
\end{equation}
with $\kappa$ still given by  \eqref{eq:kappan}. Note that the constraint $c\leq 1$ in the KPZ relations restricts the loop fugacity of the $\mathrm{FPL}(n)$ model on a bicubic map to the range $n\in [0,1]$ with $\kappa \in [6,8]$, while the complementary range $n\in (1,2)$ with $\kappa \in (4,6)$ is likely to correspond to random tree statistics.
 
A coupling between LQG and SLE with such mismatched parameters has 
yet to be described rigorously. Following \cite{D2G2}, we can simply conjecture here that for $n\in [0,1]$ the scaling limit of the $\mathrm{FPL}(n)$ model on a bicolored $p$-regular planar map with no rigid condition, will be given by $\mathrm{CLE}_\kappa$ \cite{DMS21},  with $\kappa \in [6,8]$ as in \eqref{eq:kappan}, on a $\gamma_{\scriptscriptstyle{\mathrm{L}}}$-LQG sphere with Liouville parameter  
\begin{equation}
\gamma_{\scriptscriptstyle{\mathrm{L}}}=\frac{1}{\sqrt{12}}\left(\sqrt{3\left(\kappa+\frac{16}{\kappa}\right)+22} - \sqrt{3\left(\kappa+\frac{16}{\kappa}\right)-26}\right)\ ,
\label{eq:gammaLbicubic}
\end{equation}
in agreement with conjectures proposed in \cite{BGS22,BGP22}.
\begin{figure}[htb]
  \centering
  \fig{.6}{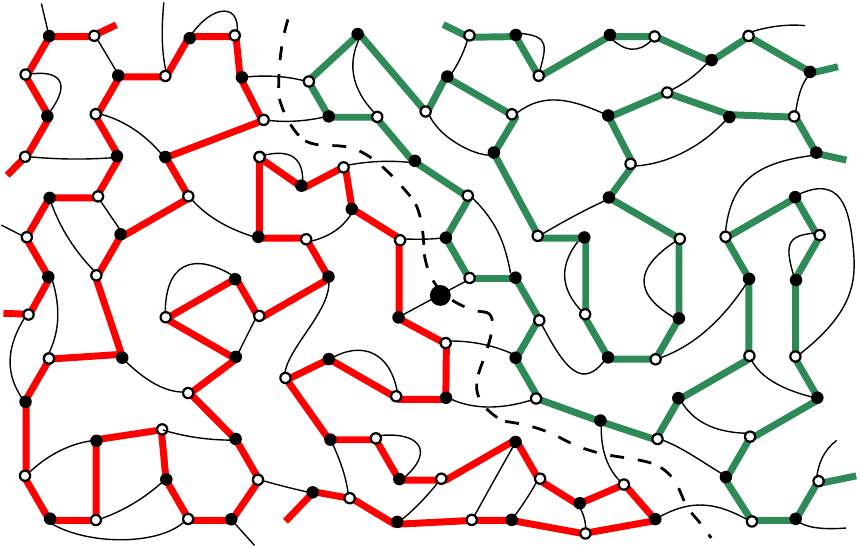}
   \caption{\small  On a random bicubic planar map with the spherical topology, the two (red and green) \emph{halves} $\mathcal C_1$ and 
   $\mathcal C_2$ of a \emph{Hamiltonian cycle} $\mathcal C=\mathcal C_1 \cup \mathcal C_2$ are separated by a (dotted) dual  
   loop $\widetilde {\mathcal C}=\mathcal C_1 \cap \mathcal C_2$ on the dual map that crosses the whole set of their nearest neighbour \emph{contact links}.  
   In the scaling limit, the random map, the fully-packed loop $\mathcal C$ and the separatrix $\widetilde {\mathcal C}$ converge (in the peanosphere topology \cite{DMS21}) to a
    $\gamma_{\scriptscriptstyle{\mathrm{L}}}$-LQG sphere decorated by a space-filling $\mathrm{SLE}_{8}$ and a whole-plane $\mathrm{SLE}_{2}$. In the case of this $(p=3)$-regular bicolored map, $c=-1$ and $\gamma_{\scriptscriptstyle{\mathrm{L}}}=\frac{1}{\sqrt{3}}\left (\sqrt{13}-1\right)$.}
  \label{fig:separatriceRandom}
\end{figure}
\subsection{Hamiltonian cycles and LQG}\label{sec:HamLQG}
The $\mathrm{FPL}(n=0)$ model on a random planar map converges to space-filling $\mathrm{SLE}_{\kappa=8}$ coupled to Liouville quantum gravity, 
 the scaling limit of a Hamiltonian cycle in the spherical topology being $\mathrm{SLE}_{8}$ decorating an independent  $\gamma_{\scriptscriptstyle{\mathrm{L}}}$-LQG sphere (for a proper definition, see  \cite{DMS21,lqg_sphere,twoperspectives}), with a Liouville parameter and a central charge depending on the choice of the map's vertex statistics.  In the case of generic (i.e., non-bicolored) cubic maps \cite{D2G2},  of bicolored maps with vertices of mixed valencies (Corollary \eqref{claim6}), and of $2q$-regular bicolored maps 
 with a local rigidity condition (Corollary\eqref{claim9}), we have from \eqref{eq:ggLk} and \eqref{eq:cfplck} for $\kappa=8$,
\begin{equation}
 \gamma_{\scriptscriptstyle{\mathrm{L}}}=\sqrt{2},\quad  \gamma=-1,\quad c=-2\ .
\label{eq:c-2}
\end{equation}
In the case of bicubic maps \cite{D2G2} or, more generally, of $p$-regular bicolored maps (Corollary \eqref{claim3}) we have from \eqref{eq:cfplck1} and \eqref{eq:gammaLbicubic} for $\kappa=8$, 
\begin{equation}
 \gamma_{\scriptscriptstyle{\mathrm{L}}}=\frac{1}{\sqrt{3}}\left (\sqrt{13}-1\right),\quad  \gamma=-\frac{1+\sqrt{13}}{6},\quad c=-1\ .
\label{eq:c-1}
\end{equation}

Let us consider the set $\widetilde{\mathcal C}=\mathcal C_1 \cap \mathcal C_2$ of contact points between the two halves of the Hamiltonian cycle $\mathcal C=\mathcal C_1 \cup \mathcal C_2$,  on a bicolored random planar map of fixed size $2N$ (see Figure \ref{fig:separatriceRandom}). 
In the thermodynamic limit $N\to \infty$, and after rescaling, this  set converges (in the peanosphere topology \cite{DMS21})  to the intersection of the two halves of an infinite $\mathrm{SLE}_8$ path, i.e., a whole-plane $\mathrm{SLE}_2$, decorating a quantum sphere of fixed  
$ \gamma_{\scriptscriptstyle{\mathrm{L}}}$-LQG area  $\mathcal A$ \cite{DMS21,lqg_sphere,twoperspectives}. 
An $\mathrm{SLE}_{\kappa=2}$ \emph{quantum length measure}  \cite{PhysRevLett.107.131305,DMS21}  based on the SLE natural parametrization \cite{ls2011natural_param}
is associated in the scaling limit with the cardinal $|\widetilde{\mathcal C}|=|\mathcal C_1 \cap \mathcal C_2|$.
Its expectation scales as
 \begin{equation}
\label{eq:contactexpquant}
{\mathbb E}_{\scriptscriptstyle{\mathrm{LQG}}}|\mathcal C_1 \cap \mathcal C_2| \asymp {\mathcal A}^\nu := {\mathcal A}^{1-\Delta_{1\cap2}}\ ,
\end{equation}
an expression entirely similar to the scaling form \eqref{eq:contactexp}, but now with a quantum exponent $\Delta_{1\cap2}:=\Delta(h_{1\cap2},c)$ given by the KPZ relation \eqref{eq:Deltahc}  in terms of $h_{1\cap2}=3/8$ \eqref{eq:hhdual}.
 Its value thus crucially depends on the central charge $c$, i.e., on the choice of vertex statistics on the bicolored map. For case  \eqref{eq:c-2}, we find
 \begin{equation}\label{eq:c-2bis}\begin{split}
&\Delta_{1\cap2}=\Delta(3/8,c=-2)=1/2\ ,\\
&\nu=1- \Delta_{1\cap2}=1/2 \ ;
\end{split}
\end{equation}
whereas in case  \eqref{eq:c-1} we predict
\begin{equation}\label{eq:c-1bis}\begin{split}
& \Delta_{1\cap2}=\Delta(3/8,c=-1)=\frac{\sqrt{11}-\sqrt{2}}{\sqrt{26}-\sqrt{2}}\ ,\\
&\nu=1- \Delta_{1\cap2}=\frac{\sqrt{26}-\sqrt{11}}{\sqrt{26}-\sqrt{2}}=0.483715\cdots \ .
\end{split}
\end{equation}
These two predictions for $\nu$ will now be tested numerically using extrapolations from exact enumerations.
\section{Numerics for long-distance contacts}
\label{sec:numcontacts}
Our Hamiltonian cycles have a marked visited edge $e$. We may thus label all the vertices by their natural order along a cycle $\mathcal C$, starting from the black vertex incident to $e$ (labelled $1$)
and ending at the white vertex incident to $e$ (labelled $2N$ if the map has size $2N$). This allows us to canonically define the two half-cycles $\mathcal C_1$ and $\mathcal C_2$ as the
parts of $\mathcal C$ containing the vertices $1$ to $N$, and $N+1$ to $2N$ respectively.
Let us denote by $k_N$ the \emph{average} number of contact links between these two halves of $\mathcal C$, see Figure ~\ref{fig:separatriceRandom}. We have
\begin{equation}
k_N =\frac{y_N}{z_N}
\label{eq:defkN}
\end{equation}
where $y_N$ denotes the partition function of Hamiltonian cycles (with a marked visited edge) of length $2N$ \emph{weighted by the number of contact links}
between their two halves.  
In the representation of Figure~\ref{fig:largearches}, this number of contacts is nothing but the number of (up or down) arches which have been opened along the first half of the 
straight line and are closed only in its second half. In the transfer matrix formalism, this number is given by the integer parts
\begin{equation}
\left\lfloor{\log_2} (\ell_u)\right\rfloor+
\left\lfloor {\log_2} (\ell_d)\right\rfloor\ 
\end{equation}
where, as in \eqref{eq:zNpair}, $|\ell_u,\ell_d\rangle$ denotes the ``middle'' state (i.e., that obtained after the action of $N$ elementary transfer matrices $T_\circ$ or $T_\bullet$).
For $N$ even, we may therefore write
\begin{equation}
\begin{split}
y_N&=\sum_{\ell_u,\ell_d}\langle1,1|( T_\circ T_\bullet)^{N/2}|\ell_u,\ell_d\rangle \big(\left\lfloor{\log_2}(\ell_u)\right\rfloor+
\left\lfloor {\log_2} (\ell_d)\right\rfloor\big) \langle \ell_u,\ell_d| ( T_\circ T_\bullet)^{N/2}  |1,1\rangle\\
&=\sum_{\ell_u,\ell_d}\left\lfloor {\log_2} (\ell_u)\right\rfloor\big(\langle \ell_u,\ell_d | ( T_\circ T_\bullet)^{N/2}  |1,1\rangle \big)^2
+ \sum_{\ell_u,\ell_d}\left\lfloor {\log_2} (\ell_d)\right\rfloor\big(\langle \ell_u,\ell_d | ( T_\circ T_\bullet)^{N/2}  |1,1\rangle \big)^2\\
&=2 \sum_{\ell_u,\ell_d}\left\lfloor {\log_2} (\ell_u)\right\rfloor\big(\langle \ell_u,\ell_d | ( T_\circ T_\bullet)^{N/2}  |1,1\rangle \big)^2\ ,
\end{split}
\end{equation}
where we used the symmetry of the problem under combined left-right reversal and black-white inversion of colors to go from the first to the second line,
as well as its up-down symmetry to go from the second to the third line. For $N$ odd, we have instead 
\begin{equation}
y_N=
2\sum_{\ell_u,\ell_d}\left\lfloor {\log_2} (\ell_u)\right\rfloor\big(\langle \ell_u,\ell_d | T_\bullet ( T_\circ T_\bullet)^{(N-1)/2}  |1,1\rangle \big)^2\ .
\end{equation}
At large $N$, we expect the asymptotic behavior
\begin{equation}
k_N  \underset{N\to \infty}{\sim} \varrho\ N^\nu
\end{equation}
with $\varrho$ depending on the bicolored map family at hand and with $\nu$ as in \eqref{eq:c-2bis} or \eqref{eq:c-1bis}. 
We expect however that the corrections to this leading behavior \emph{depend on the parity of $N$}.
This is confirmed by our numerical data: to properly estimate $\nu$  
from the sequence $(k_N)_{N\geq 1}$, we now have to split this sequence into two subsequences,  an ``even'' one  $(k_{2M})_{M\geq 1}$ and an ``odd'' one $(k_{2M-1})_{M\geq 1}$.
This leads us to define the following two independent accelerating series $(\tilde{\nu}_{2M}(s))_{M\geq 1}$ and $(\tilde{\nu}_{2M-1}(s))_{M\geq 1}$: 
\begin{equation}
\tilde{\nu}_{2M}(s)=\frac {1}{3!}(\Delta^{\!3}\, \hat{\nu})_M \quad \mathrm{with} \quad \hat{\nu}_M:=M^3 \left(M\times \mathrm{Log}\frac{k_{2M+2}+2s}{k_{2M}+2s}\right)
\label{eq:nupair}
\end{equation}
and
\begin{equation}
\tilde{\nu}_{2M-1}(s)=\frac {1}{3!}(\Delta^{\!3}\, \check{\nu})_M \quad \mathrm{with} \quad \check{\nu}_M:=M^3 \left(M\times \mathrm{Log}\frac{k_{2M+1}+2s}{k_{2M-1}+2s}\right)\ .
\label{eq:nuimpair}
\end{equation} 
Here  we introduced for future convenience an arbitrary \emph{shift} parameter $s$. Both series tends to $\nu$ at large $M$ independently of the shift $s$. The value of $s$ will eventually be
fixed numerically for each series so as to optimize the acceleration of the convergence (see below).

\medskip
It is instructive to start our analysis with the rigid $4$-regular case, for which we can write explicit expressions for $k_N$. We indeed have in this case (see Appendix ~\ref{app:r4regular})
\begin{equation}
\begin{split}
&k_{2M}+2s=\frac{2{2M\choose M}^{\!2}}{\frac{1}{2M+1}{4M \choose 2M}}+2(s-1) \ , \\ &k_{2M-1}+2s=\frac{2{2M\choose M}{2M-2\choose M-1}}{\frac{1}{2M}{4M-2 \choose 2M-1}}+2(s-1)\ .
\end{split}
\label{eq:exactkN}
\end{equation}
It is easily checked from these exact expressions that the ``even'' and ``odd'' accelerated series $(\tilde{\nu}_{2M}(s))_{M\geq 1}$ and $(\tilde{\nu}_{2M-1}(s))_{M\geq 1}$ do
converge to $\nu=1/2$ as expected, since, at large $N$, $k_{N}+2s \sim 4\sqrt{N/\pi}$ at large $N$ for any fixed $s$. In order for \eqref{eq:nupair} (resp. \eqref{eq:nuimpair}) to define a series which is effectively accelerated, i.e., for which the convergence towards $\nu$ 
is fast, it is mandatory that $\hat{\nu}_M$ (resp. $\check{\nu}_M$) have only
corrections of the form $M^{3-i}$ for integers $i\geq 1$ so that the first $3$ such corrections ($i=1,2,3$) are killed by the $3$ iterative finite difference operators $\Delta$ . 
It is easily checked from \eqref{eq:exactkN} that, in the present case, this holds only if we choose $s=1$: for $s\neq1$, $\hat{\nu}_M$ (resp. $\check{\nu}_M$) also have corrections involving half-integer powers of $M$, 
which are not killed by the finite difference operators $\Delta$, leading to a much slower convergence. Otherwise stated, the convergence to $\nu=1/2$ of 
$(\tilde{\nu}_{2M}(s))_{M\geq 1}$ (resp. $(\tilde{\nu}_{2M-1}(s))_{M\geq 1}$) is fast and reliable only if we choose $s=1$.

\begin{figure}
  \centering
  \fig{.8}{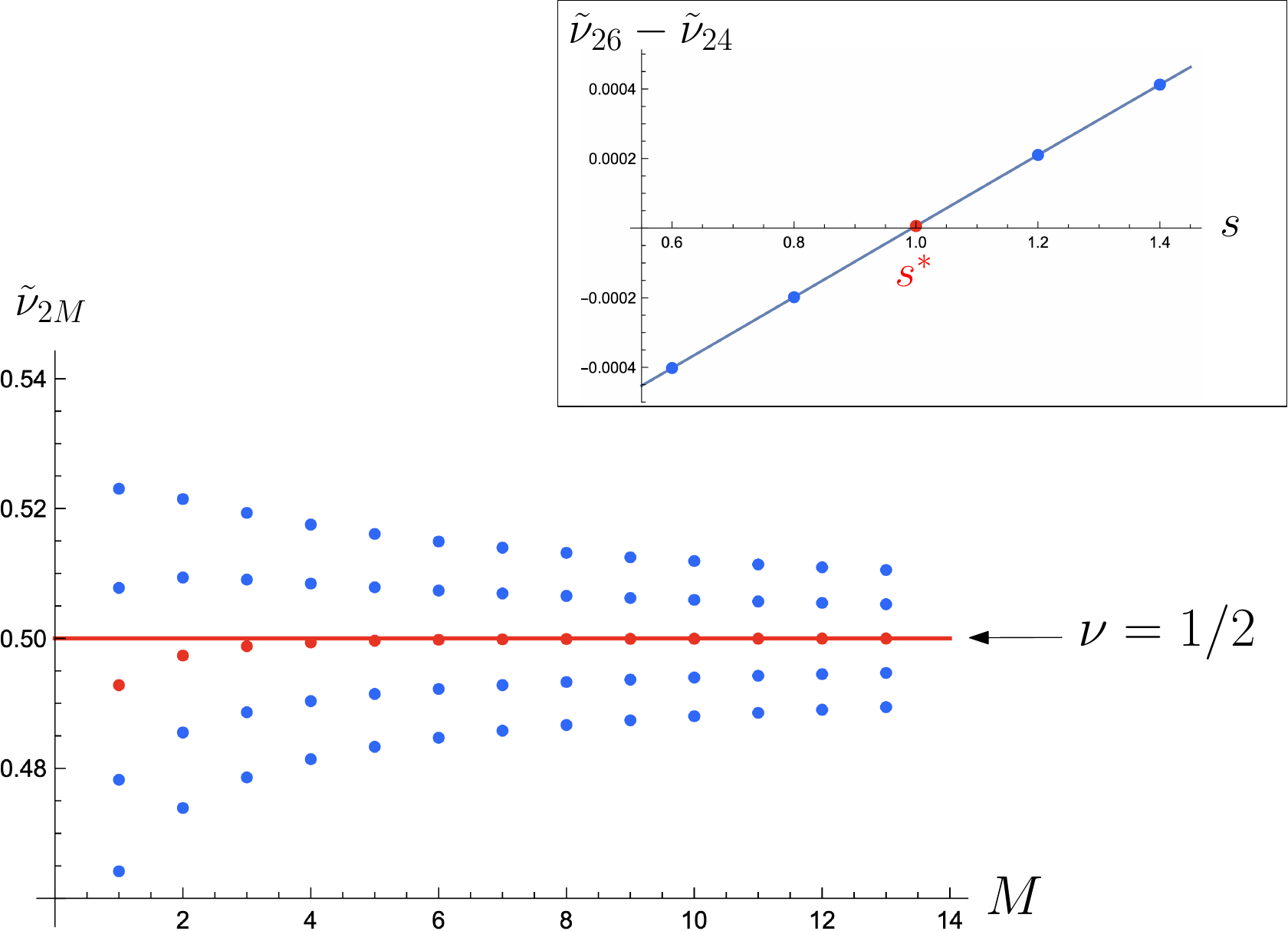}
   \caption{\small Inset: determination of the shift $s^*$ from the condition $\tilde{\nu}_{N_{\mathrm{max}}}(s^*)=\tilde{\nu}_{N_{\mathrm{max}}-2}(s^*)$ for rigid Hamiltonian cycles on $4$-regular bicolored maps
  (here with $N_{\mathrm{max}}=26$). We displayed the sequence $(\tilde{\nu}_{2M}(s))_{1\leq M \leq N_{\mathrm{max}}/2}$ for 5 different values of $s$. From top to bottom: $s=s^*-0.4$, $s=s^*-0.2$, $s=s^*$ (red), 
   $s=s^*+0.2$ and $s=s^*+0.4$. The value of $\nu$ is finally estimated from $\tilde{\nu}_{N_{\mathrm {max}}}(s^*)$ with $s^*=1.000\ ,\,\,\nu=\tilde{\nu}_{N_{\mathrm {max}}}(s^*)=0.5000$}.
  \label{fig:nupair4rigest}
\end{figure}
Suppose now that we do not know the exact expressions \eqref{eq:exactkN} and have access only to the first values of $\tilde{\nu}_{2M}(s)$ (resp. $\tilde{\nu}_{2M-1}(s)$) up to some  
finite value $N_{\mathrm{max}}=2 M_{\mathrm{max}}$ (resp. $N_{\mathrm{max}}=2 M_{\mathrm{max}}-1$). We may estimate \emph{numerically} the best value $s^*$ of $s$ by
demanding that our estimate be stabilized at $N_{\mathrm{max}}$, namely that
\begin{equation}
\tilde{\nu}_{N_{\mathrm{max}}}(s^*)=\tilde{\nu}_{N_{\mathrm{max}}-2}(s^*)\ .
\label{eq:recipe}
\end{equation}
As displayed in Figure~\ref{fig:nupair4rigest}, using as input the ``even'' accelerated series for \eqref{eq:exactkN} with $N$ up to $N_{\mathrm{max}}=26$, we obtain numerically the values
\begin{equation}
s^*=1.000\ , \qquad \nu=\tilde{\nu}_{N_{\mathrm {max}}}(s^*)=0.5000\ ,
\end{equation}
in perfect agreement with the values of $s^*$ and $\nu$ coming from the above analysis based on the exact asymptotic formulas. This therefore validates a posteriori 
our numerical recipe \eqref{eq:recipe} for the choice $s^*$ of the shift $s$. 
\begin{figure}
  \centering
  \fig{.8}{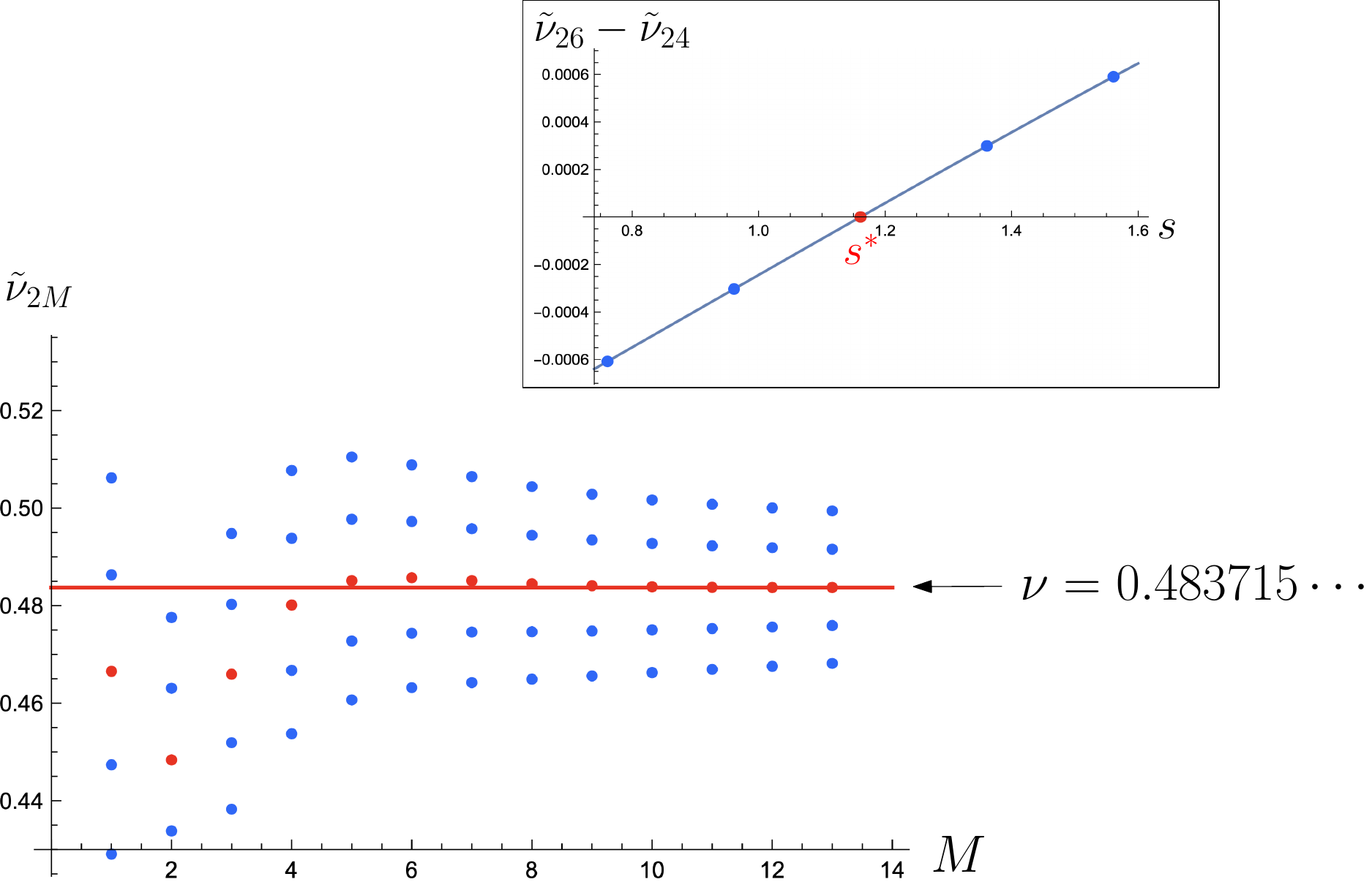}
   \caption{\small Determination of the shift $s^*$ and the exponent $\nu$ for Hamiltonian cycles on $3$-regular bicolored maps
  (with $N_{\mathrm{max}}=26$).  See caption of Figure~\ref{fig:nupair4rigest} for details.}
    \label{fig:nupair3est}
\end{figure}
\begin{figure}
  \centering
  \fig{.8}{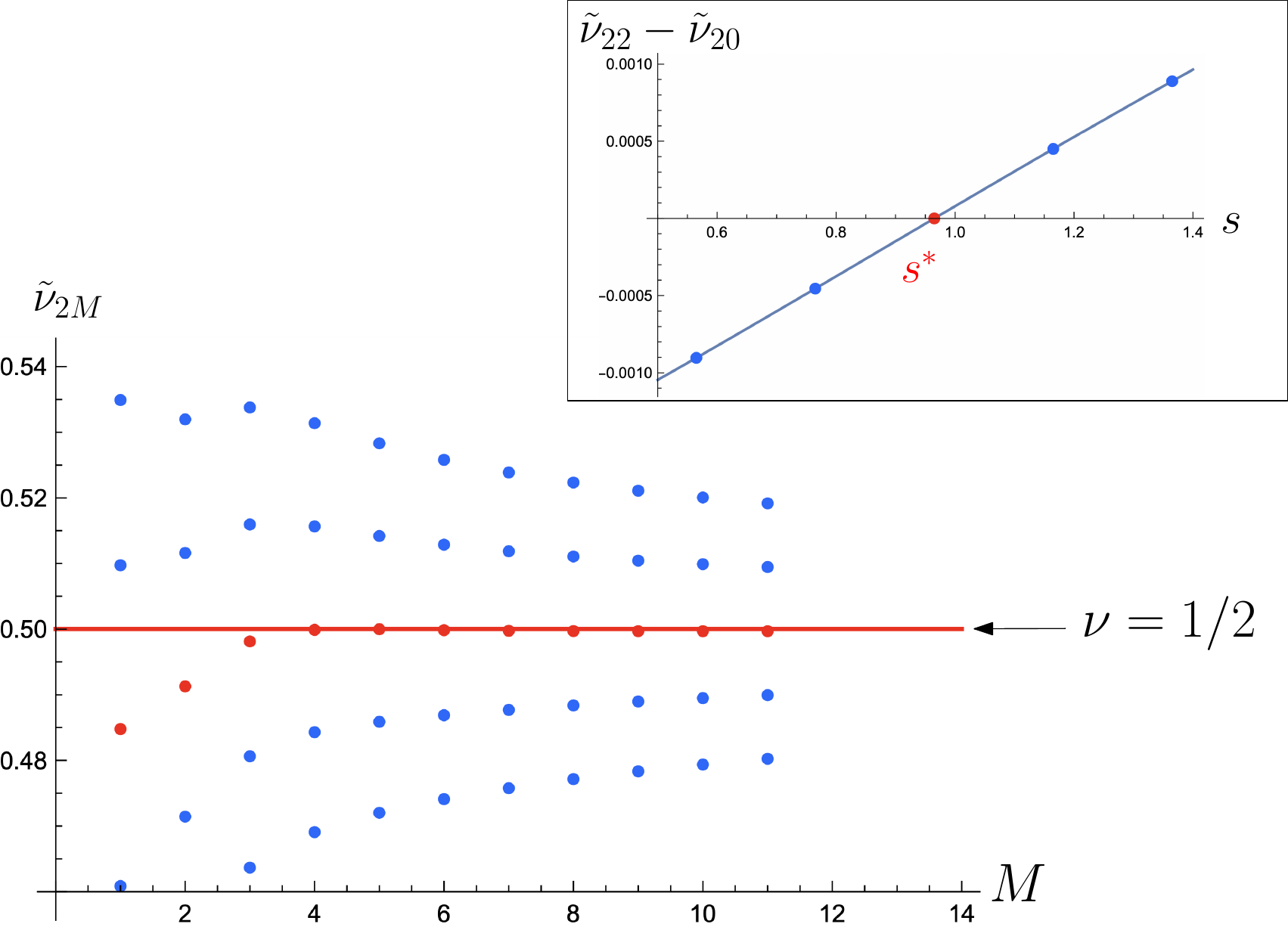}
    \caption{\small Determination of the shift $s^*$ and the exponent $\nu$ for Hamiltonian cycles on bicolored maps with mixed valencies $2$ and $3$
  (with $N_{\mathrm{max}}=22$).  See caption of Figure~\ref{fig:nupair4rigest} for details.}
  \label{fig:nupair23est}
\end{figure}

We have repeated this analysis separately with the ``even'' data and with the ``odd'' data for Hamiltonian cycles on various families of bicolored planar maps.  
For instance, Figure~\ref{fig:nupair3est} displays our results for $3$-regular bicolored maps: we get the estimates
\begin{equation}
 s^*=1.161\ , \qquad \nu=\tilde{\nu}_{N_{\mathrm {max}}}(s^*)=0.4837\ ,
 \end{equation}
hence a value of $\nu$ very close to the predicted value \eqref{eq:c-1bis}. Figure \ref{fig:nupair23est} displays similar results for maps with mixed valencies $2$ and $3$ (and
$w_2=w_3=1$), giving now $s^*=  0.965$ and $\nu= 0.4997$ very close to the predicted value $1/2$ of \eqref{eq:c-2bis}.
Table \ref{table:valnu} gives a summary of our estimates for $\nu$ for Hamiltonian cycles on six different bicolored map families and for the two parities of $N$. All the results
are in perfect agreement with the expected values.
\begin{table}[tbp]
    \centering
       \begin{tabular}{|c|c|c|c|c|c|}
        \hline
map family   &    parity of $N$  & $N_{\mathrm{max}}$ & $s^*$   &  measured $\nu=\tilde{\nu}_{N_{\mathrm {max}}}(s^*)$ & predicted $\nu$   \\
        \hline
  $3$-regular  & even &  26  & 1.161  & 0.4837  & $0.483715\cdots$ \\
  & odd &  25 &  1.185 & 0.4829 &\\
   \hline
 $4$-regular  & even  &  10  &1.008   &0.4844  & $0.483715\cdots$ \\
  & odd  & 11  & 1.054  & 0.4828 &\\
  \hline
  rigid $4$-regular   & even  & $26^{\dagger}$  & 1.000  & 0.5000 &  0.5\\
  & odd  & $25^{\dagger}$ &  1.000 & 0.5000 &  \\
  \hline
  rigid $6$-regular   & even & 22  & 0.817  & 0.5000 & 0.5  \\
   & odd  & 21  & 0.825   & 0.4999 &  \\
  \hline
 mixed valencies  & even  & 22  & 0.965  & 0.4997 & 0.5 \\  
$2$ and $3$    & odd & 21  & 0.975  & 0.4992 &  \\ 
  \hline
 mixed valencies   & even  & 8  & 0.815  & 0.4962 & 0.5  \\  
  $3$ and $4$  & odd & 7  & 0.855  & 0.4987 &  \\ 
       \hline
    \end{tabular}
   \caption{\small Estimated values of the exponent $\nu$. The value $s^*$ of the shift is determined numerically by the condition
  $\tilde{\nu}_{N_{\mathrm {max}}}(s^*)=\tilde{\nu}_{N_{\mathrm {max}}-2}(s^*)$. In the cases of mixed valencies, 
  we set $w_2=w_3=1$ (respectively $w_3=w_4=1$).\ ${}^\dagger $$[$For rigid Hamiltonian cycles on $4$-regular maps, our explicit
   expressions \eqref{eq:exactkN} allow us to take $N_{\mathrm {max}}$ arbitrarily large. The value 26 (resp. 25) was chosen for a better comparison with
   the $3$-regular case.$]$}    
   \label{table:valnu}
\end{table}
\section{Conclusion}
\label{sec:conclusion}
In this paper, we studied the statistics of Hamiltonian cycles, and more generally of fully packed loops, on various families of bicolored random planar maps
and found that the corresponding models fall into two distinct universality classes. The first, most common universality class corresponds to the coupling to
gravity of a CFT with central charge $c_{\mathrm{dense}}(n)$ as defined in \eqref{eq:valcgndense}.
This universality class is found for fully packed loops on bicolored maps with mixed valencies, for rigid fully packed loops on $2q$-regular bicolored maps, but also
for fully packed loops on non-bicolored maps (see Remark~\ref{rem:monocol}). It would also be found for non-rigid or rigid dense loops 
(i.e., O$(n)$ loops in their dense critical phase) on either bicolored or non-bicolored maps. The common feature of all these models is that they can 
be described by a single height field $\boldsymbol{\Psi}=\psi_2 \btwo$. The associated CFT on a regular lattice is that describing the dense phase
of the O$(n)$ model, with conformal dimensions which can be computed indifferently on any (hexagonal \cite{zbMATH03959008}, square \cite{PhysRevLett.58.2325}
or Manhattan \cite{dupJSP97,dupdavJSP88}) regular lattice. 
For instance, the watermelon exponent $h_\ell^{(\kappa)}$ is given 
by \eqref{eq:hkell} for any (even or odd) $\ell$, with $\kappa$ as in \eqref{eq:kappan} and its gravitational counterpart \cite{DK88,Duplantier:1989sx,KostovOn89,MR2112128} by
\begin{equation}
\Delta\big(h_\ell^{(\kappa)},c_{\mathrm{sle}}(\kappa)\big)=\frac{\ell}{4}+\frac{1}{8}(4-\kappa)\ .
\label{eq:hlkappadense}
\end{equation}

More interesting is the second universality class, corresponding to the coupling to
gravity of a CFT with central charge $c_{\mathrm{fpl}}(n)=1+c_{\mathrm{dense}}(n)$ as defined in \eqref{eq:valcgn}. This universality class is found for fully packed loops on 
$p$-regular bicolored maps for any $p\geq 3$, and corresponds to models which may now be described by a two-component height 
field $\boldsymbol{\Psi}=\psi_1\A+\psi_2 \btwo$.
In particular, we may cook up observables corresponding to (magnetic) defects (i.e., height dislocations) \emph{with a component
along the $\A$ direction}: this is the case for instance for watermelon configurations with an \emph{odd number $\ell$ of lines}. 

As already noticed in 
Section~\ref{sec:middle}, such observables are special in the sense that their conformal weights are different if we compute 
them on the (naturally bicolored) square and hexagonal regular lattices, see \eqref{eq:discry}. In this sense, universality is not
as strong for the second class (with $c=c_{\mathrm{fpl}}(n)$) as it is for the first class and only the spectrum of those
observables which \emph{do not involve the $\A$ direction} seems to be fully universal: this is in particular the case for the $2$- or $4$-line observables involved in \eqref{eq:hhdual} and associated with
the exponent $\nu$ that we considered in this paper. As for the special observables (involving the $\A$-direction), which seem to retain in the scaling limit a memory of the original lattice, one may wonder about their proper continuous description within the $\mathrm{SLE}_{\kappa}$ formalism.

When considering the watermelon configurations with an odd number of lines on $p$-regular bicolored random maps, the fact that there are two possible values for
the fully packed conformal weight $h=h^{\mathrm{fpl}(n)}_{2k-1}$ in \eqref{eq:discry}  casts some doubt on the naive use of the KPZ formula \eqref{eq:Deltahc} to get the analogue of the dense formula  \eqref{eq:hlkappadense}. Even when some choice seems ``natural'' (like for instance that of the hexagonal lattice value in \eqref{eq:discry}
when dealing 
with $3$-regular bicolored maps), it was observed in \cite{D2G2} that the associated gravitational exponent $\Delta$ is no-longer directly
related to $h$ via the KPZ formula 
\eqref{eq:Deltahc} and that some prior ``renormalization'' of the conformal weight is required. 

A subsidiary question about Hamiltonian paths on
$p$-regular bicolored maps
is therefore whether such special exponents depend on $p$, just like
they do on regular lattices with $p=3$ and $p=4$, hence lead to a weaker notion of universality. We leave this issue for a future work.

\section*{Acknowledgements} We thank P. Di Francesco for many useful discussions at the early stages of this work.
\appendix
\section{Rigid Hamiltonian cycles on $\boldsymbol{4}$-regular bicolored maps: exact enumeration formulas}
\label{app:r4regular}

The case of rigid Hamiltonian cycles on $4$-regular bicolored maps (also
called meandric systems in \cite{FT22,BGP22}) is particularly simple as we may get exact expressions for $z_N$ and $y_N$, 
hence for $k_N$ in \eqref{eq:defkN}. As already mentioned in Section~\ref{eq:exactenum}, opening the rigid cycle into a straight line of alternating black and white vertices
totally decouples the upper and lower parts, implying that
\begin{equation}
z_N=c_N^2\ ,
\label{eq:appcN}
\end{equation}
where $c_N$ enumerates non-crossing arch configurations connecting $2N$ vertices along a line \emph{on one side only}. 
Note that the fact that arches connect vertices of different colors is automatic for non-crossing arch configurations, hence we 
may forget about the colors in this particular case. As it is well known, $c_N$ is nothing but the celebrated Catalan number 
\begin{equation}
c_N=\frac{1}{N+1}{2N\choose N}\ ,
\end{equation}
in agreement with \eqref{eq:fusscatalan} for $q=2$. 
\begin{figure}[h]
  \centering
  \fig{.8}{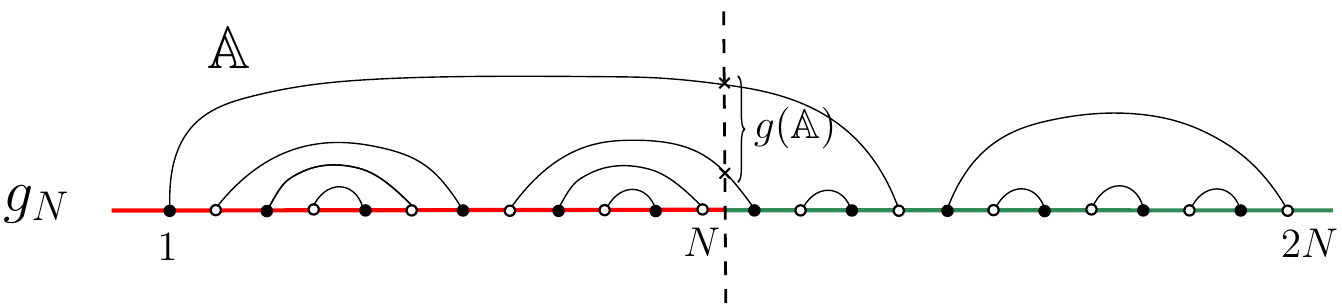}
   \caption{\small An arch configuration $\mathbb{A}$ contributing to $g_N$ (here with $N=12$) and the number $g(\mathbb{A})$ of arches passing above the middle 
point (here $g(\mathbb{A})$=2), whose parity is the same as that of $N$. }
  \label{fig:gN}
\end{figure}
Let us now discuss the quantity $y_N$. The decoupling of the upper and lower parts (together with the up-down symmetry) implies that
\begin{equation}
y_N=2 g_N c_N
\label{eq:appgN}
\end{equation}
where $g_N$ enumerates arch configurations $\mathbb{A}$ connecting $2N$ vertices along a line \emph{on one side only}, weighted by the number $g(\mathbb{A})$ of arches passing 
above the middle point of the straight line (i.e., the middle point of the edge connecting the $N$-th to the $(N+1)$-th vertex), see Figure~\ref{fig:gN}. Let us first assume that $N$ is even and write $N=2M$. 
This implies that $g(\mathbb{A})$ is even too. More precisely, for $0\leq p\leq M$, those arch configurations  $\mathbb{A}$  for which $g(\mathbb{A})=2p$ are enumerated by\footnote{In the Dyck path representation of non-crossing arch systems \cite{stanley2015}, these configurations correspond to pairs made of (i) a path of length $2M$ from height $0$ to height $2p$ (hence with $M+p$ up-steps) and (ii) 
a complementary path of length $2M$ from height $2p$ to height $0$ (hence with $M+p$ down steps), both paths having only non-negative heights.}
\begin{equation}
\left({2M\choose M+p}-{2M\choose M+p+1}\right)^{\!2}= \left( \frac{2p+1}{M+p+1}{2M\choose M+p}\right)^{\!2}\ .
\end{equation} 
This yields
\begin{equation}
\begin{split}
g_N&=\sum_{p=0}^M \left( \frac{2p+1}{M+p+1}{2M\choose M+p}\right)^{\!2} (2p) \\
&= \sum_{p=0}^M \left( \frac{2p+1}{M+p+1}{2M\choose M+p}\right)^{\!2} (2p+s) - s\, c_N\ ,\\
\end{split}
\label{eq:gNexp}
\end{equation} 
where we used the sum rule $\sum\limits_{p=0}^M \left( \frac{2p+1}{M+p+1}{2M\choose M+p}\right)^{\!2}=c_N$ for the total number of arch configurations.
Noting that
\begin{equation}
\left( \frac{2p+1}{M+p+1}{2M\choose M+p}\right)^{\!2} (2p+1)=\Delta_p \left(-\frac{(M+2p^2)}{M}{2M\choose M+p}^{\!2} \right)
\end{equation}
where $\Delta_p$ is the forward finite difference operator in $p$, we see that the sum in the second line of \eqref{eq:gNexp} is telescopic for the choice $s=1$.

We eventually end up with
\begin{equation}
g_N={2M\choose M}^{\!2}-c_{N}\quad \mathrm{for}\ N=2M\ ,
\label{eq:gNq2val}
\end{equation}
and 
\begin{equation}
k_{N}=\frac{y_N}{z_N}=\frac{2g_N}{c_N}=\frac{2{2M\choose M}^{\!2}}{\frac{1}{2M+1}{4M \choose 2M}}-2\quad \mathrm{for}\ N=2M\ .
\label{eq:kNpair}
\end{equation}
If we now assume that $N$ is odd, a similar calculation leads to
\begin{equation}
k_{N}=\frac{2{2M\choose M}{2M-2\choose M-1}}{\frac{1}{2M}{4M-2 \choose 2M-1}}-2\quad \mathrm{for}\ N=2M-1\ .
\label{eq:kNimpair}
\end{equation}
Eqs. \eqref{eq:kNpair} and \eqref{eq:kNimpair} lead to the desired formulas \eqref{eq:exactkN}.

\begin{table}
\section{Numerical data}
\label{app:numerics}
  \centering
  \begin{tabular}{|rrr|}
    \hline
    $N$ &                          $z_N$ &                         $y_N/2$ \\
    \hline
     1 &                               2 &                               1 \\
     2 &                               8 &                               4 \\
     3 &                              40 &                              28 \\
     4 &                             228 &                             182 \\
     5 &                            1424 &                            1376 \\
     6 &                            9520 &                           10256 \\
     7 &                           67064 &                           82256 \\
     8 &                          492292 &                          657258 \\
     9 &                         3735112 &                         5483168 \\
    10 &                        29114128 &                        45720644 \\
    11 &                       232077344 &                       392225248 \\
    12 &                      1885195276 &                      3367237302 \\
    13 &                     15562235264 &                     29496561288 \\
    14 &                    130263211680 &                    258689070208 \\
    15 &                   1103650297320 &                   2303183835424 \\
    16 &                   9450760284100 &                  20532423715862 \\
    17 &                  81696139565864 &                 185194267822952 \\
    18 &                 712188311673280 &                1672505538588120 \\
    19 &                6255662512111248 &               15246126785026456 \\
    20 &               55324571848957688 &              139146249302900840 \\
    21 &              492328039660580784 &             1279654964632731016 \\
    22 &             4406003100524940624 &            11781309072368013800 \\
    23 &            39635193868649858744 &           109156077594746888256 \\
    24 &           358245485706959890508 &          1012371771569816836390 \\
    25 &          3252243000921333423544 &          9439721149094472748640 \\
    26 &         29644552626822516031040 &         88100169337671128409824 \\
    27 &        271230872346635464906816 &        826012547472307809557896 \\
    28 &       2490299924154166673782584 &       7751024033279177862804200 \\
    29 &      22939294579586403144527440 &      73022459752163336202562352 \\
    30 &     211949268051816569236796848 &     688468559155925660846596544 \\
    31 &    1963919128426791258770276024 &    6513579576440364032532422976 \\
    32 &   18246482008315207478524287044 &   61667572983605062268400200798 \\
    33 &  169953210523325203868381657400 &  585630198026539853341680121888 \\
    34 & 1586759491069775179474823509344 & 5565011094981145493511752402704 \\
    \hline
  \end{tabular}
  \caption{Values of $z_N$ (sequence A116456 in OEIS \cite{OEIS}) and $y_N$ for Hamiltonian cycles on bicolored 3-regular planar maps.}
  \label{table:3regular}
\end{table}
\begin{table}
  \centering
  \begin{tabular}{|rrr|}
    \hline
    $N$ &                      $z_N$ &                      $y_N/2$ \\
    \hline
     1 &                           3 &                            3 \\
     2 &                          34 &                           34 \\
     3 &                         583 &                          797 \\
     4 &                       12370 &                        18962 \\
     5 &                      299310 &                       541218 \\
     6 &                     7914962 &                     15658990 \\
     7 &                   223112249 &                    492077299 \\
     8 &                  6599227954 &                  15610597634 \\
     9 &                202656932134 &                 519177791710 \\
    10 &               6413548643796 &               17387351622688 \\
    11 &             208040580206216 &              600403799410348 \\
    12 &            6888733433298402 &            20842604582620710 \\
    13 &          232117149975205154 &           739230697828101014 \\
    14 &         7939206408814949506 &         26327452538168278582 \\
    15 &       275098365065617821621 &        952653521434740072227 \\
    16 &      9641385973628938712306 &      34586535913246138331782 \\
    17 &    341313811643888153301006 &    1271131209796113395573406 \\
    18 &  12191280053256623302185704 &   46844535638524226902706228 \\
    19 & 438954593201892408379178942 & 1743184882186466069552567270 \\
    \hline
  \end{tabular}
  \caption{Values of $z_N$ and $y_N$ for Hamiltonian cycles on bicolored 4-regular planar maps.}
  \label{table:4regular}
\end{table}
\begin{table}
  \centering
  \begin{tabular}{|rrrr|}
    \hline
    $N$ &                5-regular &               6-regular &             7-regular \\
    \hline
     1 &                         4 &                       5 &                     6 \\
     2 &                       104 &                     259 &                   560 \\
     3 &                      4640 &                   25094 &                104024 \\
     4 &                    266084 &                 3192155 &              25715048 \\
     5 &                  17669760 &               474183765 &            7462790096 \\
     6 &                1292292432 &             77907665840 &         2401948332096 \\
     7 &              101201942512 &          13740308705438 &       831180015105160 \\
     8 &             8340015146964 &        2554205527336363 &    303462839364701128 \\
     9 &           714995787362600 &      494475099243189329 & 115462177891927344416 \\
    10 &         63259444105430512 &    98867302126812855515 &                       \\
    11 &       5742719613679409832 & 20294465583102673352590 &                       \\
    12 &     532599319939460085760 &                         &                       \\
    13 &   50295898068432583524224 &                         &                       \\
    14 & 4823733144104904305892304 &                         &                       \\
    \hline
  \end{tabular}
\caption{Values of $z_N$ for Hamiltonian cycles on bicolored 5-regular, 6-regular and 7-regular planar maps.}
  \label{table:567regular}
  
\end{table}
\begin{table}
  \centering
  \begin{tabular}{|rrr|}
    \hline
    $N$ &                             $z_N$ &                            $y_N/2$ \\
    \hline
     1 &                                  3 &                                  1 \\
     2 &                                 17 &                                  6 \\
     3 &                                125 &                                 67 \\
     4 &                               1077 &                                676 \\
     5 &                              10335 &                               8047 \\
     6 &                             107151 &                              93898 \\
     7 &                            1176999 &                            1184387 \\
     8 &                           13518677 &                           14869772 \\
     9 &                          160872323 &                          195389839 \\
    10 &                         1970329025 &                         2566924518 \\
    11 &                        24715305741 &                        34751956495 \\
    12 &                       316322082895 &                       471076136766 \\
    13 &                      4118646279649 &                      6523535179149 \\
    14 &                     54428554176853 &                     90491263299716 \\
    15 &                    728662270487961 &                   1275474547319661 \\
    16 &                   9866887839946229 &                  18009066127518820 \\
    17 &                 134967673222112567 &                 257454410282564295 \\
    18 &                1862969746410518745 &                3686602712849035850 \\
    19 &               25924506623086706277 &               53316166797618448047 \\
    20 &              363415643231059957421 &              772238458092154850980 \\
    21 &             5128518034166712107763 &            11276238109326334073237 \\
    22 &            72814980427431398768943 &           164883291621449041519854 \\
    23 &          1039603583945087464438759 &          2427283275342458095362671 \\
    24 &         14918925552410770296750503 &         35777211288494249743148062 \\
    25 &        215108422239328159518817305 &        530360761101151938386907819 \\
    26 &       3115114976238433506239203399 &       7870933845679033785904203612 \\
    27 &      45295058700528813260672278919 &     117382878931669305354337886003 \\
    28 &     661097024940535265310437647345 &    1752373351490083766149516091464 \\
    29 &    9682937008170057158267261746831 &   26271697196196181749006295843637 \\
    30 &  142290916972981046011294091297071 &  394231951670046541461277392969298 \\
    31 & 2097420196208084754056265923088015 & 5937785334543529526068890061788573 \\
    \hline
  \end{tabular}
  \caption{Values of $z_N$ and $y_N$ for Hamiltonian cycles on bicolored planar maps with mixed valencies 2 and 3 (with $w_2=w_3=1$).}
  \label{table:2-3-mixed}
\end{table}
\begin{table}
  \centering
    \begin{tabular}{|rr|rr|}
    \hline
    $N$ &                       $z_N$  & $N$ &                       $z_N$ \\
    \hline
     1 &                            4 & 10 &               24584694155437 \\
     2 &                           47 & 11 &              930530200722914 \\
     3 &                          872 & 12 &            36039351335158162 \\
     4 &                        20579 & 13 &          1423250588260168692 \\
     5 &                       562346 & 14 &         57153474076536198864 \\
     6 &                     16959202 & 15 &       2328611379453123805998 \\
     7 &                    549029380 & 16 &      96085895789053111221723 \\
     8 &                  18750074923 & 17 &    4009433404474389044318028 \\
     9 &                 667653126308 & 18 &  168976691280496979237329801 \\
    \hline
  \end{tabular}
  \caption{Values of $z_N$ for Hamiltonian cycles on bicolored planar maps with mixed valencies 2 and 4 (with $w_2=w_4=1$).}
  \label{table:2-4-mixed}
\end{table}
\begin{table}
  \centering
  \begin{tabular}{|rrr|}
    \hline
    $N$ &                      $z_N$ &                      $y_N/2$ \\
    \hline
     1 &                           5 &                            4 \\
     2 &                          98 &                           80 \\
     3 &                        3089 &                         3572 \\
     4 &                      124622 &                       163552 \\
     5 &                     5844034 &                      9159648 \\
     6 &                   303138220 &                    522941716 \\
     7 &                 16901630655 &                  32699927584 \\
     8 &                994850903414 &                2071909682642 \\
     9 &              61080867353216 &              138275419169022 \\
    10 &            3878907227559258 &             9315849112395598 \\
    11 &          253224873797465540 &           649064156160267680 \\
    12 &        16915976848381443504 &         45541980819371884184 \\
    13 &      1152241256370476649482 &       3271499179479967664002 \\
    14 &     79806203708523623827632 &     236287877905404626333174 \\
    15 &   5608021949255349143950993 &   17365297252695426225180534 \\
    16 & 399095475044872817013511142 & 1281725711268335772862571494 \\
    \hline
  \end{tabular}
  \caption{Values of $z_N$ and $y_N$ for Hamiltonian cycles on bicolored planar maps with mixed valencies 3 and 4 (with $w_3=w_4=1$).}
  \label{table:3-4-mixed}
\end{table}

\begin{table}
As already seen in Section \ref{sec:rigid} (Eqs. \eqref{eq:zc2} \eqref{eq:fusscatalan}) and in Appendix~\ref{app:r4regular} when $q=2$, using the arch representation such as that of Figure~\ref{fig:largearches}
in the case of \emph{rigid} Hamiltonian cycles on $2q$-regular bicolored planar maps for arbitrary $q\geq 2$ leads to a complete decoupling between
the upper and lower arch configurations. This implies the following the two identities (extending \eqref{eq:appcN} and \eqref{eq:appgN}):
\begin{equation}
z_N= c_N^2\quad \mathrm{with}\ 
c_N=\frac{1}{(q-1)N+1}{q\, N\choose N}
\label{eq:appzN}
\end{equation}
and 
\begin{equation}
y_N=2 g_N c_N\ ,
\label{eq:apphN}
\end{equation}
where $g_N$ enumerates arch configurations \emph{on one side only}, weighted by the number of arches passing 
above the middle point of the straight line, see Figure~\ref{fig:gN} when $q=2$.
We have no exact expression for $g_N$ for arbitrary $q\geq 3$ (which would generalize \eqref{eq:gNq2val}). 
The following table gives the first values of $g_N$ in the case $q=3$, from which we can get $y_N$ via \eqref{eq:apphN}. \hfill

\medskip 
  \centering
  \begin{tabular}{|rr|rr|}
    \hline
    $N$ &                  $g_N$ & $N$ &                  $g_N$ \\
    \hline
     1 &                       2 &  16 &            429765359266 \\
     2 &                       6 &  17 &           2747996363358 \\
     3 &                      32 &  18 &          17558452105246 \\
     4 &                     162 &  19 &         112880676289328 \\
     5 &                     930 &  20 &         725294746632006 \\
     6 &                    5260 &  21 &        4683479629941570 \\
     7 &                   31432 &  22 &       30229921171815208 \\
     8 &                  186606 &  23 &      195925602453080976 \\
     9 &                 1142582 &  24 &     1269396826660493508 \\
    10 &                 6971466 &  25 &     8252873289420323592 \\
    11 &                43385904 &  26 &    53640502233395278680 \\
    12 &               269429292 &  27 &   349671835181599650032 \\
    13 &              1696338360 &  28 &  2278921678933838458548 \\
    14 &             10665144516 &  29 & 14890267787292439785072 \\
    15 &             67735129000 &  30 & 97273104239590589753820 \\
    \hline
  \end{tabular}
  \caption{Values of $g_N$ (such that $y_N=2 g_N c_N$ with $c_N$ as in \eqref{eq:appzN}) for rigid Hamiltonian cycles on $6$-regular bicolored planar maps (i.e., $q=3$).}
  \label{table:6rigid}
\end{table}

\newpage
\bibliographystyle{unsrt}
\bibliography{HamiltBicolor}
\end{document}